\documentclass[twocolumn]{jpsj2} 
\title{Time-dependent DMRG Study on Quantum Dot under a Finite Bias Voltage}

\author{Shunsuke Kirino, Tatsuya Fujii, Jize Zhao and Kazuo Ueda}

\inst{
 Institute for Solid State Physics, University of Tokyo, Kashiwa
 277-8581, Japan
}

\abst{%
 Resonant tunneling through quantum dot under a finite bias
 voltage at zero temperature is investigated by using the adaptive
 time-dependent density matrix renormalization group(TdDMRG) method.
 Quantum dot is modeled by the Anderson Hamiltonian with the 1-D
 nearest-neighbor tight-binding leads. 
 Initially the ground state wave function is calculated with the usual
 DMRG method. 
 Then the time evolution of the wave function due to the slowly
 changing bias voltage between the two leads is calculated by using the
 TdDMRG technique. 
 Even though the system size is finite, the expectation values of
 current operator show steady-like behavior for a finite time interval,
 in which the system is expected to resemble the real nonequilibrium
 steady state of the infinitely long system.
 We show that from the time intervals one can obtain quantitatively correct
 results for differential conductance in a wide range of bias voltage.
 Finally we observe an anomalous behavior in the expectation value of
 the double occupation operator at the dot
 $\langle n_{\uparrow} n_{\downarrow} \rangle$ as a function of bias voltage.
}

\kword{
quantum dot, nonequilibrium, Kondo effect, adaptive time-dependent DMRG}

\usepackage{graphicx}	
\begin{document}
\maketitle

\section{Introduction}
Recently it has been established that the Kondo effect plays an
important role in transport properties of quantum dot systems
at low temperatures \cite{linear_1, linear_2, exp_1, exp_2, exp_3}.
A new feature of the Kondo effect in quantum dot systems in comparison
with the traditional magnetic impurity problems is that nonequilibrium
steady state is realized under a finite bias voltage. 

In order to study the properties of the steady states theoretically, 
analyses based on Keldysh formalism are often employed.
There are two different types of approaches to quantum dot out of
equilibrium:
approaches from the noninteracting limit such as the perturbation theory
with respect to the Coulomb interaction $U$ \cite{2nd_order, Fujii_1, Fujii_2}, 
and approaches from the strong coupling limit such as 
noncrossing approximation \cite{NCA},
real-time diagrammatic formulation \cite{Konig}
and renormalization group method applied to the s-d model
\cite{Rosch_1,Rosch_2}. 

In the absence of a magnetic field, 
an anomalous peak structure other than the zero bias peak was predicted
in differential conductance by 
the $4$th order perturbation theory with respect to the Coulomb
interaction $U$ \cite{Fujii_1}.
Recently this new peak was obtained also in the functional renormalization group
result \cite{Gezzi}.
The peak appears when the bias voltage exceeds the Kondo temperature,
$eV \sim k_B T_K$, and it is a challenging problem to describe the
crossover theoretically.

In finite magnetic fields, it is established both experimentally and theoretically
that the zero bias peak splits into two and the peaks are located
where the bias voltage is equal to the Zeeman splitting,
$eV= \pm g \mu_B h$ \cite{exp_1,exp_2,Amasha,NCA,Fujii_2}.

However, the parameter range where each type of theories can be applied is
limited.
For example, the perturbation theory is applicable in
a relatively weak coupling regime \cite{Fujii_2} and
the noncrossing approximation \cite{NCA} and the renormalization group
method \cite{Rosch_1,Rosch_2} are confined to the strong coupling regime.
In the present situation a better treatment is required, that covers all 
the parameter space of Coulomb interaction, bias voltage and magnetic
field. 

Apart from Keldysh formalism, it seems that there are only few
numerical studies on quantum dot under a finite bias voltage.
This is largely due to a lack of reliable numerical techniques for the
nonequilibrium problems of the mesoscopic systems.

One possible numerical approach to the quantum dot out of equilibrium
is the adaptive time-dependent density matrix renormalization group (TdDMRG)
method \cite{tdDMRG_1,tdDMRG_2}, by which one can accurately calculate time
evolution of wave function of one dimensional system.
In the previous TdDMRG study on a quantum dot system \cite{dot+tdDMRG}
the results were practically limited in the linear response regime.
So the $V$ dependence of the physical quantities in the nonequilibrium
steady state, such as the differential conductance, were not discussed
in their report.

In this paper we investigate the zero temperature transport properties
of the quantum dot in a wide range of bias votage by the TdDMRG method.
We demonstrate that, in a limited time interval, the current obtained by the
finite system calculation provides us
accurate information about the one in the steady state of the infinitely
long system. 
Then we show that the $V$ dependence of the physical quantities can be
reliably obtained from the time intervals.

\section{Model and Calculation}

\subsection{Hamiltonian}
We consider one single level quantum dot with two leads.
This system is described by the Anderson model.
The universal feature of the Kondo effect allows us to
use 1-D nearest-neighbor tight-binding model for the lead parts.
In this paper we concentrate on the symmetric case for simplicity.
Then the model to be studied is described as
\begin{align}
 \label{Hamiltonian_all}
 H=
 &-t \sum_{i < -1, 0 < i} \sum_{\sigma} ( c_{i
 \sigma}^{\dagger} c_{i+1 \sigma} + h.c.)    \notag \\
 &- t' \sum_{\sigma} \left[ \left( c_{-1 \sigma}^{\dagger} c_{0 \sigma}
 + h.c. \right) + \left( c_{0 \sigma}^{\dagger} c_{1 \sigma}
 + h.c. \right) \right]    \notag \\
 &+ \sum_{\sigma} \left( -\frac{U}{2} - \tilde h \sigma \right) c_{0 \sigma}^{\dagger} c_{0
 \sigma} + U c_{0 \uparrow}^{\dagger} c_{0 \uparrow} c_{0
 \downarrow}^{\dagger} c_{0 \downarrow},
\end{align}
where the quantum dot is placed at the $0$th site.
In the above equation $t$ is the hopping amplitude in the leads,
$t'$ the hopping amplitude between the leads and the dot,
$U$ the Coulomb energy and
$\tilde h \sigma$ the Zeeman energy.
The one particle energy at the dot site is set to
$-U/2$ under the symmetric condition.

\subsection{Process to reach steady state}
In Keldysh formalism, one starts with an equilibrium state of the
unperturbed Hamiltonian. 
One then turns on the perturbation term adiabatically, and
gets a nonequilibrium steady state after sufficiently long time.

There are several choices of the perturbation terms to be turned on.
It is expected that the system reachs the same steady state independent
of the choices.
For the present problem, Anderson model out of equilibrium, there are
two frequently used options:
one is to change the hopping amplitudes $t'$ adiabatically, keeping the
chemical potentials of the two leads constant in time.
The other is to set $t'$ constant and change the chemical potentials.
In Keldysh-based theories the former one is often used.
Note that the interaction term is also turned on simultaneously.
Here we take the latter choice by adding the time-dependent bias term,
\begin{align}
 H_{\mathrm{bias}}(\tau) = \frac{eV}{2} \theta(\tau) \sum_{\sigma}
 \left[ \sum_{i<0} c_{i \sigma}^{\dagger} c_{i \sigma}
 - \sum_{i>0} c_{i \sigma}^{\dagger} c_{i \sigma} \right],
\label{bias_term}
\end{align}
to the Hamiltonian eq.\eqref{Hamiltonian_all}.
$\tau$ represents the real time variable and
$\theta(\tau)$ is a smoothed step function which we define as
\begin{align}
 \theta(\tau) = \frac{1}{ \exp{\left(\frac{\tau_0 - \tau}{\tau_1}\right)} +1}.
\end{align}

The reason for our choice of the time dependence of the Hamiltonian is that
when $t'=0$ the number of nonzero eigenvalues of the reduced density matrix
of the lead parts is only one, so the usual DMRG procedure to obtain
the optimal basis set cannot be used directly.

\subsection{Calculation of current}
Based on Keldysh formalism for the quantum
dot out of equilibrium, the steady current at $T=0$ is calculated by
\cite{2nd_order}
\begin{align}
\label{nonequilibrium_current}
 J(V)=\frac{e}{\hbar}\sum_{\sigma} \int_{\mu_R}^{\mu_L} d \omega \frac{\Gamma_{L
 \sigma} \Gamma_{R \sigma}}{\Gamma_{L \sigma} + \Gamma_{R \sigma}}
 \rho_{\sigma}(\omega),
\end{align}
where $\mu_L,\mu_R$ are the chemical potentials of the two leads,
$\Gamma_{L,R \sigma}$ the resonance widths due to the mixing between the
dot and the leads, and $\rho_{\sigma}(\omega)$ the spectral function at the dot site. 
For the present model
\begin{align}
 \Gamma_{L,R \sigma}(\omega) = \pi t'^2 D_{L,R \sigma}(\omega),
\end{align}
where $D_{L,R \sigma}(\omega)$ is the local density of states at the
edge of the lead.
The bias voltage $V$ appears in $\Gamma_{L,R}, \rho_{\sigma}(\omega)$
and as $\mu_L - \mu_R = eV$ in eq.\eqref{nonequilibrium_current}.
The essential part to compute eq.\eqref{nonequilibrium_current}
is reduced to the calculation of $\rho_{\sigma}(\omega)$.

On the other hand, in the TdDMRG calculation we simply get the current
by taking the expectation value of the current operator with the wave
function at each time. 
We may define two current operators: one flows from the left lead to the
dot and the other from the dot to the right lead,
\begin{align}
& J_L(\tau) = \langle \psi(\tau) |
 \frac{ie t'}{2 \hbar} \sum_{\sigma}
  (c_{0 \sigma}^{\dagger} c_{-1 \sigma} - h.c.)
 | \psi(\tau)
\label{J_L} \rangle,  \\ 
& J_R(\tau) = \langle \psi(\tau) |
 \frac{ie t'}{2 \hbar} \sum_{\sigma}
  (c_{1 \sigma}^{\dagger} c_{0 \sigma} - h.c.)
 | \psi(\tau) \rangle.
\label{J_R}
\end{align}
In the TdDMRG the essential part to obtain the current
becomes the calculation of $| \psi(\tau) \rangle$.
This can be done by the TdDMRG method.

In the noninteracting case, we can easily compute
$\rho_{\sigma}(\omega)$ exactly and obtain $J(V)$ for infinitely long leads.
Alternatively, when $U=0$ we can also use exact diagonalization for the
Hamiltonians without the bias term
and calculate $| \psi(\tau) \rangle$ exactly by assuming sudden
switching-on of the bias voltage. 
We use these results to compare with the TdDMRG results in later sections.

\section{Adaptive Time-dependent DMRG Method \label{Sec:TdDMRG}}
Here we briefly explain the numerical technique we use.
The basic idea is to combine the Suzuki-Trotter decomposition of the 
time evolution operator and the DMRG finite-system algorithm with 
the wave function prediction method \cite{White_Prediction}.

Our Hamiltonian eqs.\eqref{Hamiltonian_all} and \eqref{bias_term} can be 
written as the sum of the $i$th bond Hamiltonian $h_{i}(\tau)$.
Then the time evolution operator is found to be
\begin{align}
 U(\tau_0+\Delta \tau,\tau_0)
 &= \mathrm{T} \exp{\left(-\frac{i}{\hbar} \int_{\tau_0}^{\tau_0+\Delta \tau} 
 \sum_{i} h_{i}(\tau) d \tau \right)}
    \notag \\
 &\simeq e^{-\frac{i}{\hbar} \frac{\Delta \tau}{2} h_1(\tau_0+\frac{\Delta
 \tau}{2})} \cdots 
 e^{-\frac{i}{\hbar} \frac{\Delta \tau}{2} h_{L-1}(\tau_0+\frac{\Delta \tau}{2})}   \notag \\
 & \times e^{-\frac{i}{\hbar} \frac{\Delta \tau}{2} h_{L-1}(\tau_0+\frac{\Delta \tau}{2})}
 \cdots e^{-\frac{i}{\hbar} \frac{\Delta \tau}{2} h_1(\tau_0+\frac{\Delta \tau}{2})},
\label{S-T_2}
\end{align}
by using the 2nd order Suzuki-Trotter decomposition of the T-ordered exponential
\cite{S-T_decomposition}.
Then the time evolution operator is well approximated as the product of the
local time evolution operators $U_{i}^{\mathrm{local}}$.

The ground state wave function obtained from the usual DMRG calculation
for open boundary condition has the form of 
\begin{align}
 | \psi \rangle =
  \sum_{\alpha_l \sigma_{l+1} \sigma_{l+2} \beta_{l+3}}
  \psi_{\alpha_l \sigma_{l+1} \sigma_{l+2} \beta_{l+3}} 
  |\alpha_l \rangle |\sigma_{l+1} \rangle |\sigma_{l+2} \rangle
  |\beta_{l+3} \rangle,
\end{align}
where 
$|\alpha_l \rangle, |\sigma_{l+1} \rangle, |\sigma_{l+2} \rangle, |\beta_{l+3} \rangle$
are the DMRG basis set of the left block (which consists of $l$ sites), left site,
right site and right block, respectively. 
The operation of the local time evolution operator at the sites $l+1$ and
$l+2$ can be calculated without any error as
\begin{align}
& \left(U_{l+1}^{\mathrm{local}} \psi \right)_{\alpha_l \sigma_{l+1}
 \sigma_{l+2} \beta_{l+3}}   \notag \\
& \hspace{0.3cm}
 = \sum_{\sigma'_{l+1} \sigma'_{l+2}}
 \left(U_{l+1}^{\mathrm{local}}\right)_{\sigma_{l+1} \sigma_{l+2};
 \sigma'_{l+1} \sigma'_{l+2}}
 \psi_{\alpha_l \sigma'_{l+1} \sigma'_{l+2} \beta_{l+3}}.
\end{align}
Then the wave function prediction method is used to move to the next
configuration of the finite system algorithm, that is, the number of
the basis to describe the system block and the one additional site is
truncated to a fixed number $m$.

Performing this procedure at every step of the DMRG finite system algorithm, 
the full operation of the decomposed time evolution operator eq.\eqref{S-T_2} 
on the wave function can be done in one full sweep, keeping the basis
set optimal.

The main sources of errors involved in this method are the Trotter error,
which comes from the Suzuki-Trotter decomposition of the time evolution operator, 
and the DMRG truncation error, which originates from the reduction
of the number of block basis to $m$ in the wave function prediction method.
As the calculation proceeds the errors acummulate step by step and the TdDMRG result
gradually loses its accuracy.

The TdDMRG results shown in this paper are obtained with $\Delta \tau =0.05 \hbar /t$.

\section{Time Dependence of the Current \label{Sec:4}}
Since we are interested in the properties of the nonequilibrium steady
states of the infinitely long systems,
the most important requirement for the present TdDMRG calculation is to
realize the steady states numerically.
But it is obvious that the steady states cannot be realized in a finite
system in a rigorous sense.

In this section we show that one can find steady-like behaviors 
in finite systems that mimic behaviors of steady states of infinite
systems
and discuss how to obtain the information on the steady states
from the TdDMRG calculations.

\subsection{Oscillations in $J_L(\tau)$ and $J_R(\tau)$}
The total particle number and $\sum_{i} S_i^z$ are the conserved quantities of
the Hamiltonian eq\eqref{Hamiltonian_all} and eq.\eqref{bias_term}.
Thus it is sufficient to consider only the states which belong to the 
subspace with fixed particle number and $\sum_{i} S_i^z$.
In this paper we concentrate on the half-filled and $\sum_{i} S_i^z=0$ 
subspace, where the Hamiltonian is the most symmetric and the Kondo
effect is the most significant.
Then the system length $L$ should be taken as an even number and the
lengths of the left and right leads cannot be the same.
We set the right lead is longer than the left lead by one site.

\begin{figure}[t]
 \begin{center}
  \includegraphics[width=4.5cm,angle=270]{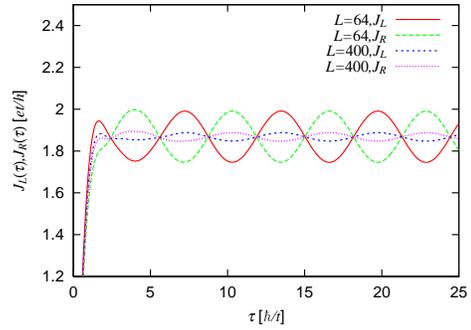}
  \caption{(Color online) $J_L(\tau)$ and $J_R(\tau)$ with different system sizes
  ($L=64$ and $L=400$) obtained by the exact diagonalization.
  The parameters used are $t'/t=0.6, U/t=0, eV/t=1.0$ and $\tilde h/t=0$.
  \label{Fig:oscillation}}
 \end{center}
\end{figure}%
$J_L(\tau)$ and $J_R(\tau)$ defined in eqs.\eqref{J_L} and \eqref{J_R}
should be the same in steady states in infinite system.
However 
for a finite system these currents show oscillations \cite{Schmitteckert} as in 
Fig.\ref{Fig:oscillation}.
It can be seen that the amplitude of the oscillation is smaller for
$L=400$ than for $L=64$ and therefore we conclude that this
oscillation disappears in the limit $L\rightarrow \infty$. 
For the present model under the symmetric condition $J_L(\tau)$ and $J_R(\tau)$
oscillate in opposite phase. 
Consequently we use the averaged current
\begin{align}
 J(\tau) = \frac{1}{2} \left(J_L(\tau) + J_R(\tau) \right),
\end{align}
in order to remove the oscillations.

\subsection{Behavior of the current in long time scale \label{Sec:reflection}}
\begin{figure}[t]
 \begin{center}
  \includegraphics[width=4.5cm,angle=270]{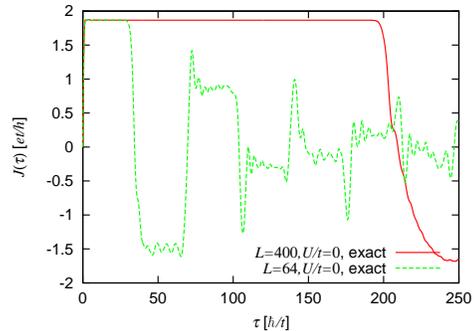}
  \caption{(Color online) $J(\tau)$ for $L=64$ and $L=400$ obtained by the exact diagonalization.
  The parameters used are $L=400, t'/t=0.6, U/t=0, eV/t=1.0$ and
  $\tilde h/t=0$.
  \label{Fig:long_time}}
 \end{center}
\end{figure}

The long time behavior of $J(\tau)$ is dominated mainly by the system
size $L$.
In Fig.\ref{Fig:long_time} we see an obvious
difference between $L=64$ and $L=400$ results after $\tau \sim 30 \hbar /t$.
This can be explained as follows:
electron wave packets driven by the bias voltage will eventually arrive at
the right edge of the system, and be reflected without any loss of energy.
After some time the wave packets will get back to the center of the
system and make negative contributions to $J(\tau)$.
$J(\tau)$ for $L=64$ after $\tau \sim 30 \hbar/t$ results from 
a complicated superposition of the wave packets going left and right.

Because this reflection at the edges of the system is an artifact of
the finite size calculation, we do not consider
$J(\tau)$ after this effect appears.

\subsection{Switching-on of bias voltage and response}
A characteristic energy of the system in equilibrium is the renormalized
resonance 
width $\tilde \Gamma \equiv \Gamma / \tilde \chi_{\uparrow \uparrow}$,
where $\tilde \chi_{\uparrow \uparrow}$ is defined as \cite{Hewson}
\begin{align}
 \tilde \chi_{\sigma \sigma'} \equiv \delta_{\sigma \sigma'} -
 \frac{\partial \Sigma^r_{\mathrm{eq} \sigma}(0)}{\partial (\tilde h \sigma')}
 \bigg| _{\tilde h \sigma' =0},
\end{align}
using the retarded self energy for $V=0$, $\Sigma^r_{\mathrm{eq} \sigma}(\omega)$.
$\tilde \chi_{\sigma \sigma'}$ in equilibrium state can be calculated by
the Bethe Ansatz solution.
$\tilde \Gamma$ is a monotonically decreasing function of $U$ 
and becomes the Kondo temperature $T_K$ times a constant in
the strong coupling limit. 

Response of the system to the switching-on of bias term takes place in
the time scale of the order of $\hbar /\tilde \Gamma$.
When the time variation of the bias term is faster compared to
$\hbar /\tilde \Gamma$, $J(\tau)$ shows an overshoot behavior followed
by a damping oscillation.
Therefore in order to get smoother results for $J(\tau)$, it is preferable to use
slower switching-on of the bias term.
On the other hand, the result in long time scale becomes unreliable
because of the reflections at the edge of the system
and the acummulation of errors.

In an actual experimental situation the resonance width $\Gamma$ is
much smaller than the band width $4t$.
However we take rather large value of $\Gamma$, with the aim to
accelerate the response of the system and to lead the system to reach
the steady state before the calculation loses its reliability.
We fix $t'/t$ as $0.6$, which means $\Gamma(\omega=0) |_{V=0} = 0.72 t$.
In addition we set $\tau_0=3\hbar/t, \tau_1=\hbar/t$.
We will see later that we can obtain reasonable results by using these values
of $t'$ and the parameters in the smoothed step function.

\subsection{Quasi-steady state in finite system \label{Sec:quasi-steady}}
\begin{figure}[t]
 \begin{center}
  \includegraphics[width=4.5cm,angle=270]{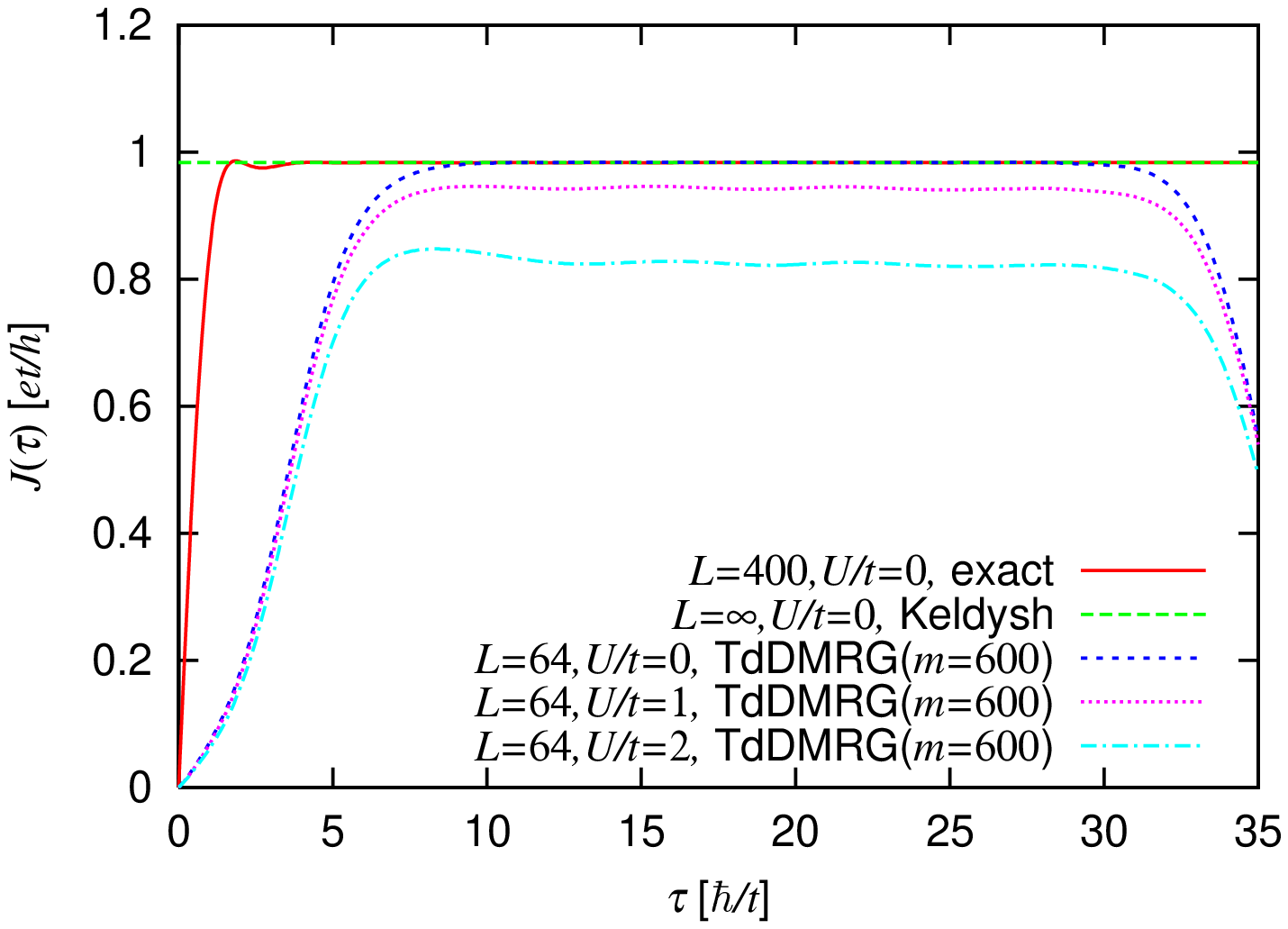}
  \includegraphics[width=4.5cm,angle=270]{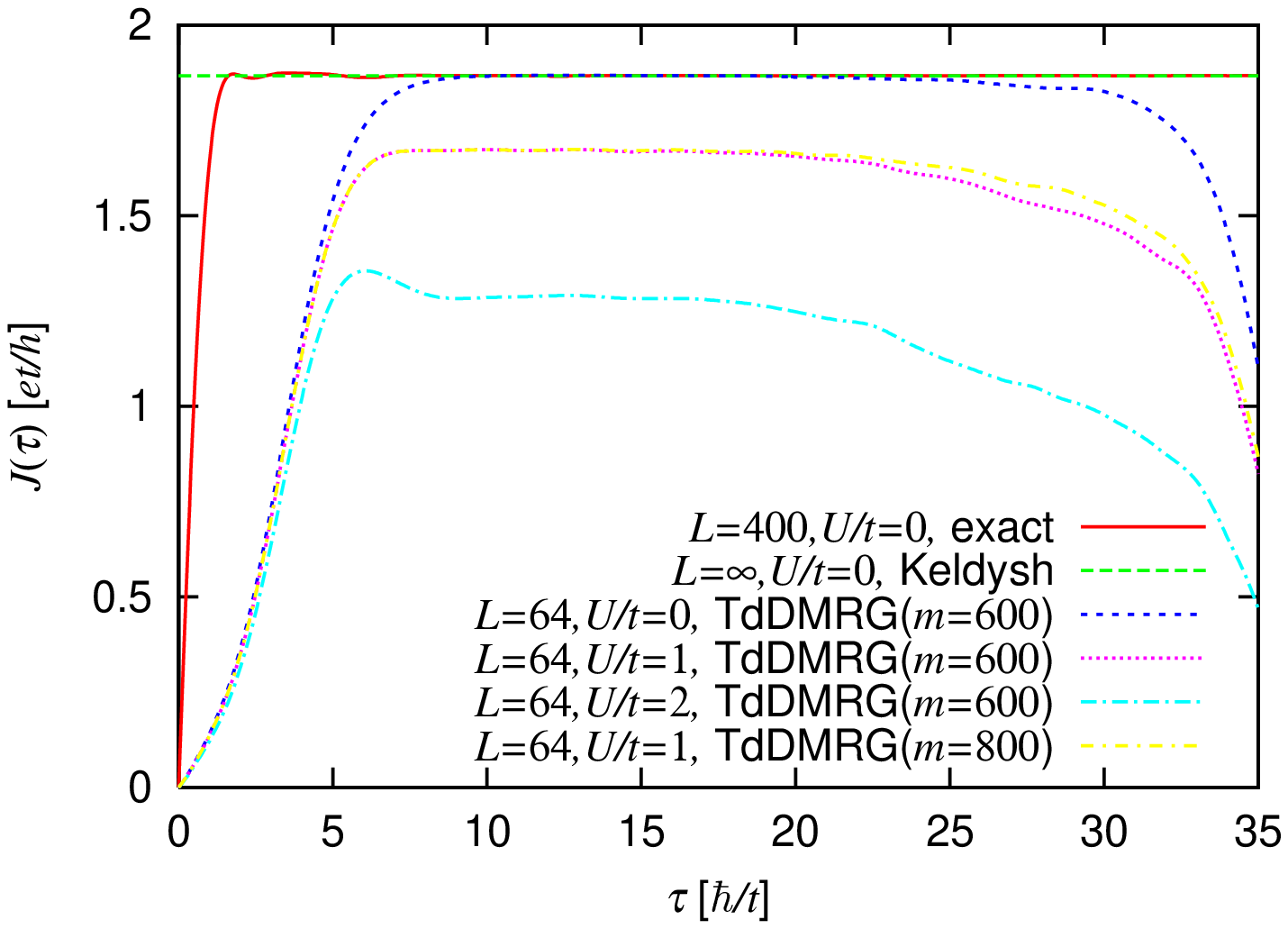}
  \includegraphics[width=4.5cm,angle=270]{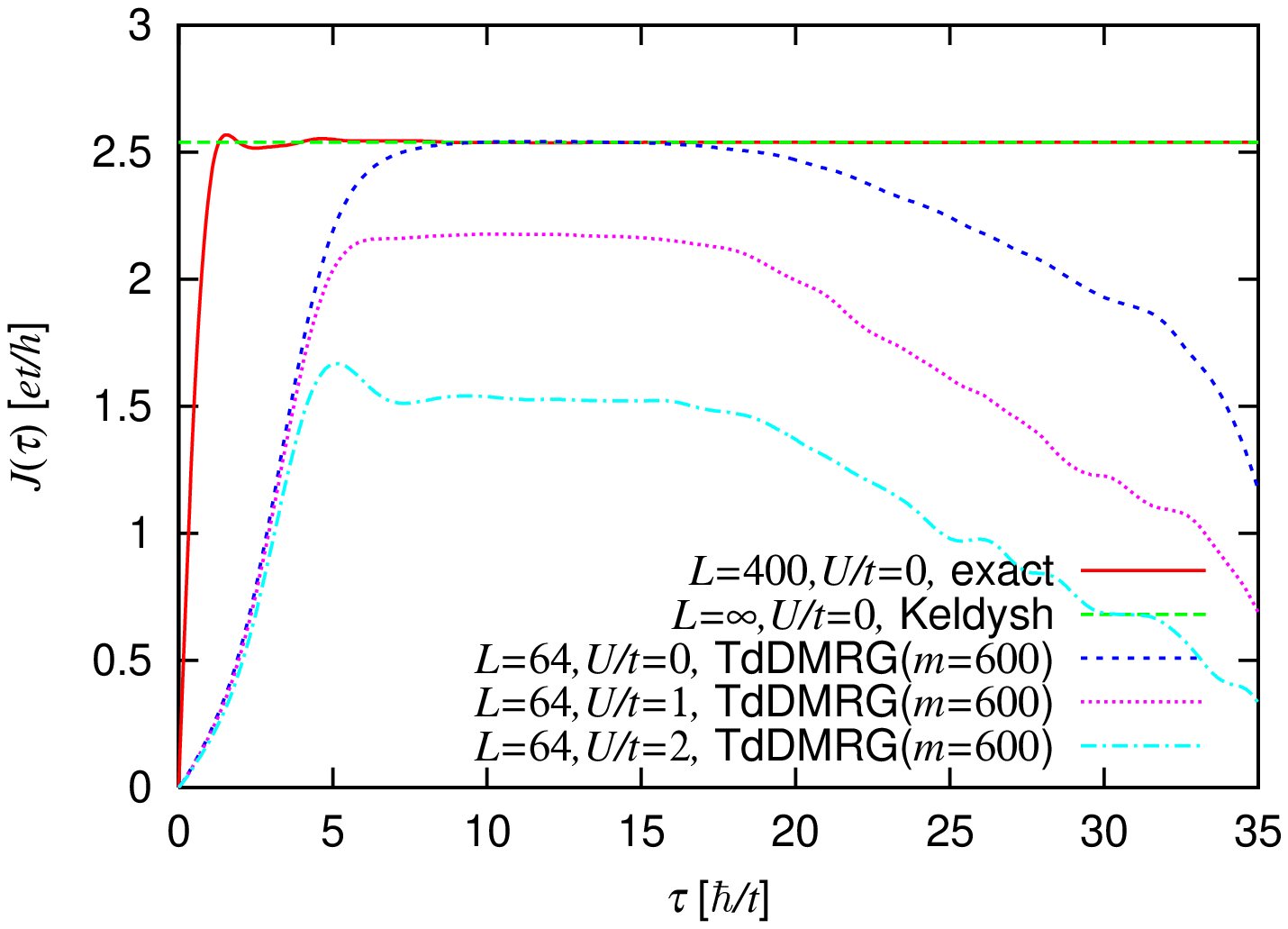}
  \includegraphics[width=4.5cm,angle=270]{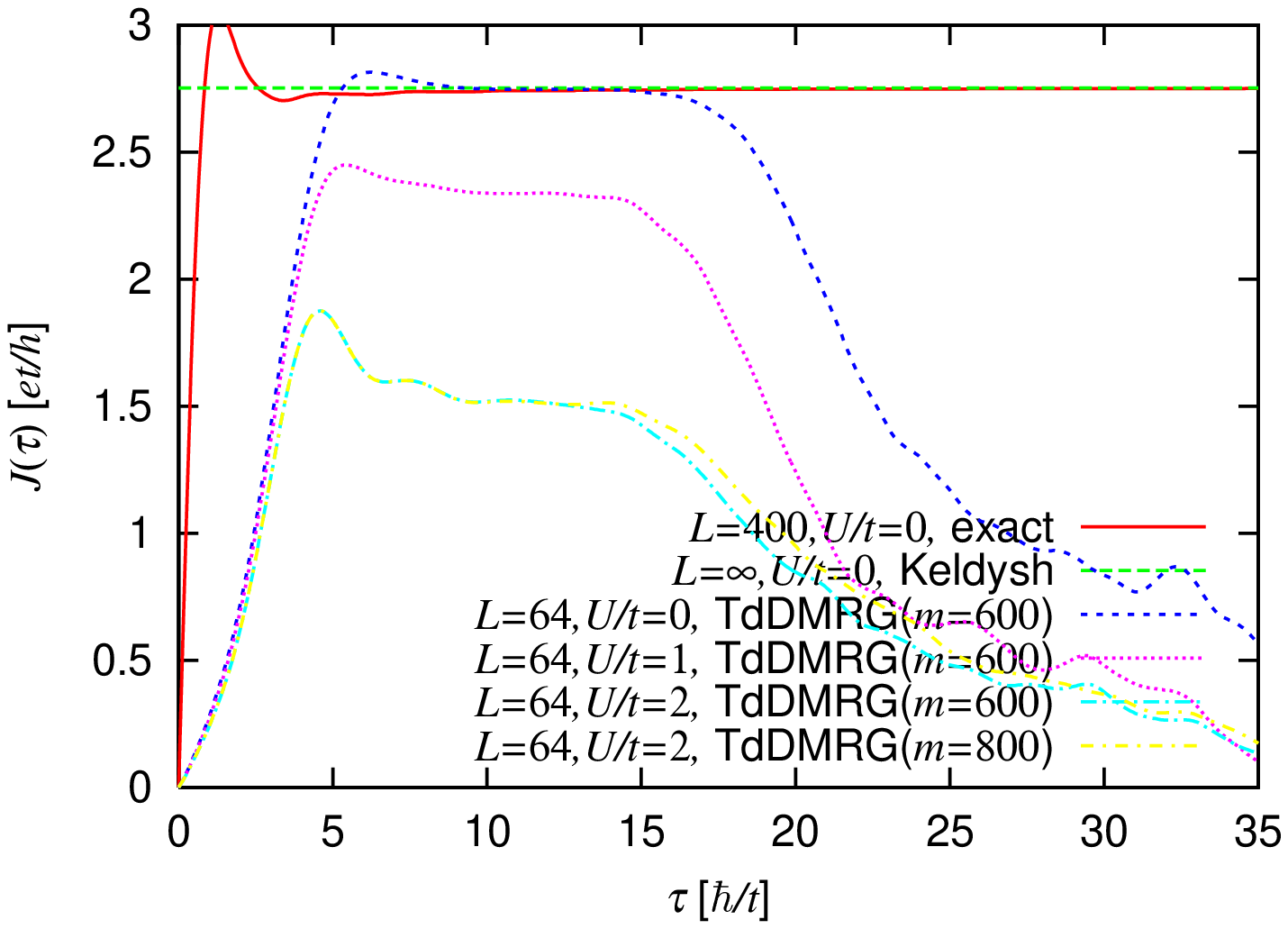}
  \caption{(Color online) $J(\tau)$ obtained by analytic calculation ($L=\infty, U=0$), exact
  diagonalization ($L=400, U=0$) and TdDMRG ($L=64, U/t=0,1,2$).
  The steady-like behaviors can be seen in all the TdDMRG results shown
  here except for the one $U/t=2,eV/t=2,m=600$.
  The parameters used are $t'/t=0.6, \tilde h/t=0$ and $eV/t=0.5,1.0,1.5,2.0$
  from the top.
  \label{Fig:comparison}}
 \end{center}
\end{figure}

As already stated, in the noninteracting case we can use three different
methods to calculate current: integration of
eq.\eqref{nonequilibrium_current}, exact diagonalization and TdDMRG method. 
In Fig.\ref{Fig:comparison} we compare the results obtained from these methods.
For $U/t=0$ the three results show good agreement in the flat region:
$8 \hbar/t < \tau < 30 \hbar/t$ for $eV/t=0.5$,
$8 \hbar/t < \tau < 23 \hbar/t$ for $eV/t=1.0$,
$8 \hbar/t < \tau < 16 \hbar/t$ for $eV/t=1.5$ and
$8 \hbar/t < \tau < 15 \hbar/t$ for $eV/t=2.0$.
Note that the system sizes and the processes to bring the system to the
steady state are quite different.
Nevertheless the currents in the flat region show a good coincidence.
Thus we can conclude that the system reaches similar steady state
independent of the choice of the perturbation term.
Moreover, because $L=\infty$ for the analytic calculation and $L=64$ for
the TdDMRG, we may also conclude that the system size dependence of the
steady current is small
under the symmetric condition.
Out of the symmetric condition this system size dependence is
considerably large \cite{Schmitteckert}, and one has to employ the
finite size scaling to obtain results for the infinite system.

Even for $U/t=1$ or $2$ we can get the well defined plateaus of $J(\tau)$ in
Fig.\ref{Fig:comparison}.
For $U/t=2, eV/t=2$ the TdDMRG result with $m=600$ does not stay at a
particular value but with $m=800$ the plateau appears from
$\tau \simeq 9 \hbar/t$ to $\tau \simeq 14 \hbar/t$.
In these plateau regions the system is expected to simulate the
nonequilibrium steady states of the infinitely long system, just as in
the $U=0$ case.

Hereafter we call the system in the flat region where $J(\tau)$ stays
almost constant within $\pm 1 \%$ as a quasi-steady state in finite system.
Then it is natural to define the steady current as an average value over
the quasi-steady state region. 
We will see later that this definition gives consistent results with the
Keldysh results for not only $U=0$ but also $U \neq 0$ at least in the
low bias regime, see \S \ref{Sec:current_0}.

It is worth noting that the overshoots discussed in the previous
subsection are observed in Fig.\ref{Fig:comparison}.
We see that the overshoots become bigger for larger $U$ and $V$.
This is due to the small value of $\tilde \Gamma$, the energy scale of
the system, relative to the time variation of the bias term.
These overshoots are expected to disappear if we use more slowly-changing
bias voltage, in other words, if we use larger values of $\tau_0, \tau_1$.
But in this paper we do not try to fine-tune the optimal choice to reduce the
overshoots, because the quasi-steady states are well-defined in the
parameter range we have studied.

\subsection{Decay of quasi-steady state}
As we have seen, we cannot realize the true steady states by the TdDMRG method.
We can only realize the quasi-steady states that have finite lifetimes.

There are several factors that determine the end of the quasi-steady state.
One is the reflection of the electron wave packets at the edges of the system,
as discussed in \S \ref{Sec:reflection}.
Another is due to the change of the population of electrons in each
lead.
When the number of electrons in the left lead becomes too few, 
the current cannot keep on flowing steadily.
The third one is the acummulation of the truncation error, see
\S \ref{Sec:TdDMRG}.

The larger bias voltage $V$ means the more current flows and
the larger change occurs in the wave function by the time evolution.
Then the second and third factors become important when $V$ is large.
Therefore, the larger $V$ we use, the shorter the lifetime of the
quasi-steady state becomes.
From Fig.\ref{Fig:comparison} we can confirm that the decay happens earlier
with increasing bias voltage.
Moreover we can see that by increasing $m$, in other words by reducing
the truncation error, the decay of the quasi-steady state can be delayed
as in the results for $U/t=1, eV/t=1$ and $U/t=2, eV/t=2$.

Note also that from the results $U/t=1, eV/t=1, m=600$ and $m=800$, the $m$
dependence of the current in the plateau region is small.
Thus we can expect that once the clearly recognizable quasi-steady
state is realized the steady current obtained is very close to the
exact value for $m\rightarrow \infty$.

Eventually it is seen that the quasi-steady states are well realized for 
$0\leq U/t \leq 2, 0.5 \leq eV/t \leq 2$ by using the parameters chosen.

\subsection{Boundary conditions}
It is well known that the Kondo cloud spreads as $U$ increases.
In other words, the lower energy excitations of the leads become
important with increasing $U$.

From this reason one has to employ sufficiently large system
to describe the Kondo effect successfully by DMRG.
However it is numerically difficult because 
a large system requires a large 
number of basis to be kept for a given accuracy.
Then the calculation for larger $U$ takes much more time.

\begin{figure}[t]
 \begin{center}
  \includegraphics[width=4.5cm,angle=270]{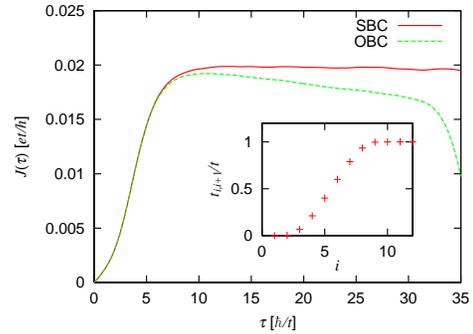}
  \caption{(Color online) TdDMRG results for $J(\tau)$ with different boundary conditions.
  It can be seen from the OBC result that the system size is not sufficient
  to describe the Kondo effect.
  Obviously the SBC result is much better than the OBC one in the sense
  of finding the quasi-steady state.
  The parameters used are $L=64, m=600, t'/t=0.6, U/t=2.0, eV/t=0.01$
  and $\tilde h/t=0$.
  Inset: The hopping amplitudes with SBC near the left edge of the
  system obtained by eq.\eqref{SBC_hopping}.
  \label{Fig:boundary_conditions}}
 \end{center}
\end{figure}

For the present calculation, the low bias regime is difficult to treat
because the low energy excitations are mainly involved in the transport, and
thus the size effect is strong.
We can improve effeciency of the calculation in the parameter range by using a
different boundary condition other than the usual open boundary
condition (OBC) \cite{SBC,DBC,dot+tdDMRG}. 
The idea is to make the energy scale near the boundary of the lead small
by reducing the hopping amplitude exponentially towards the edges of the
system, thus including the lower energy excitations.
There are of course several possible ways of reducing the hopping
parameters.
In this paper we use the smooth boundary condition \cite{SBC} (SBC)
by setting the $10$ hopping parameters from the edge as
\begin{align}
 t_{i,i+1}/t = \frac{1}{2} \left(1 +
 \tanh\frac{x_i-1/2}{x_i(1-x_i)}\right)
\label{SBC_hopping}
\end{align}
where $x_i = (i-1/2)/10$, as shown in the inset of
Fig.\ref{Fig:boundary_conditions} (left side).
In Fig.\ref{Fig:boundary_conditions} we see dramatic improvement
in the realization of the quasi-steady state.
While the OBC result does not show the steady-like behavior,
we clearly observe a quasi-steady state for the SBC result.
This implies that our choice of the hopping parameters
(eq.\eqref{SBC_hopping}) works well for the parameters in
Fig.\ref{Fig:boundary_conditions}.

It should be mentioned that there are some shortcomings of using the SBC.
First is that the SBC slows down the convergence of the
diagonalization step in the usual DMRG method.
This is due to the facts that the diagonalization of a large sparse
matrix is done by a power method such as Lanczos or Davidson method,
and that the SBC bring the first excited energy closer to the ground state
energy.
The second is that in the SBC results the truncation errors become more
significant than in the OBC results.
So we use the SBC for $eV/t \leq 0.3$, otherwise we use the OBC.
By setting the boundary conditions in this way, we can obtain the
quasi-steady states for $0 \leq U/t \leq 2, \, 0 \leq eV/t \leq 2$.

\subsection{$1/L$ correction for $\tilde h \neq 0$ \label{Sec:1/L}}
\begin{figure}[t]
 \begin{center}
  \includegraphics[width=4.5cm,angle=270]{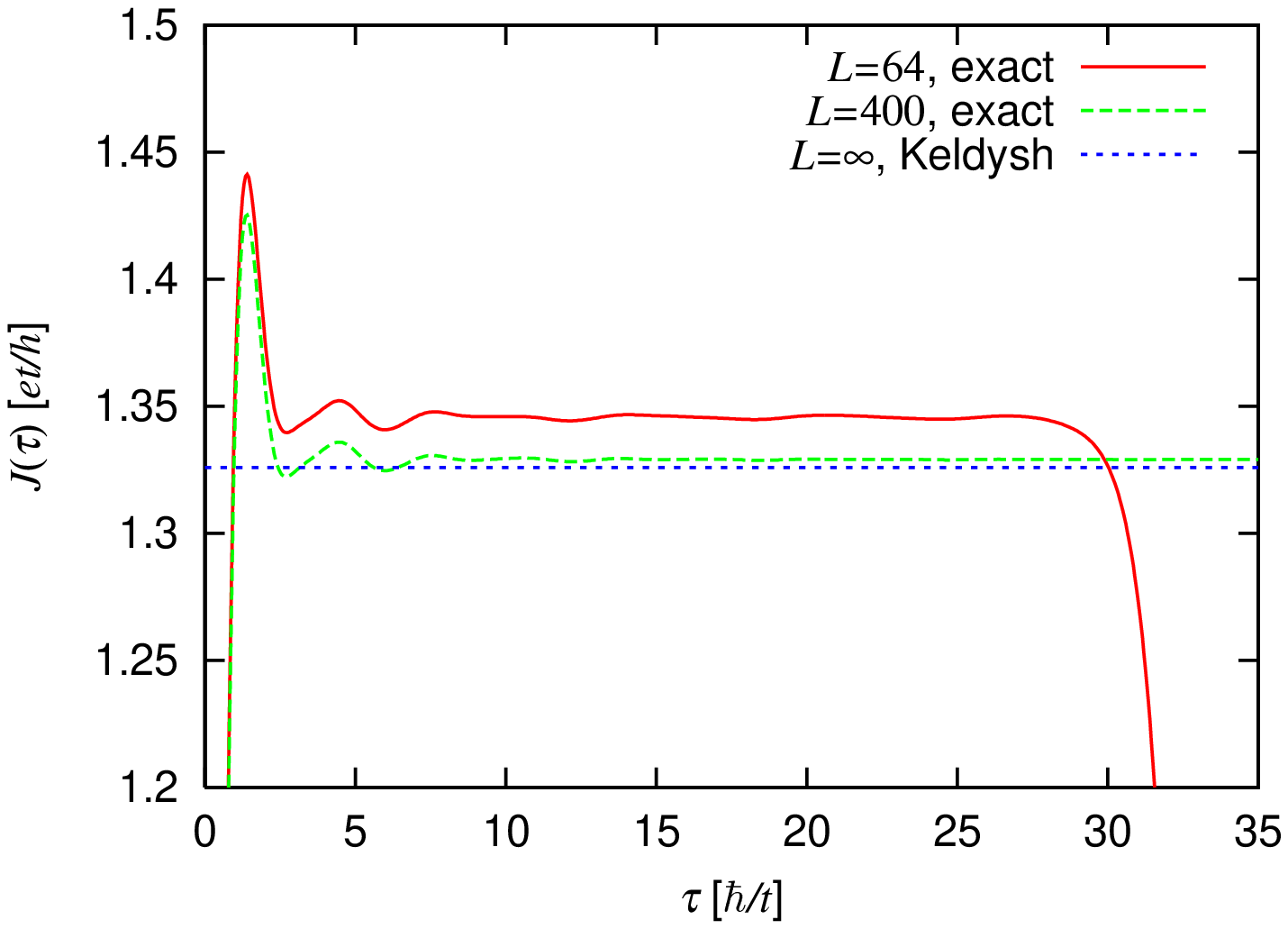}
  \includegraphics[width=4.5cm,angle=270]{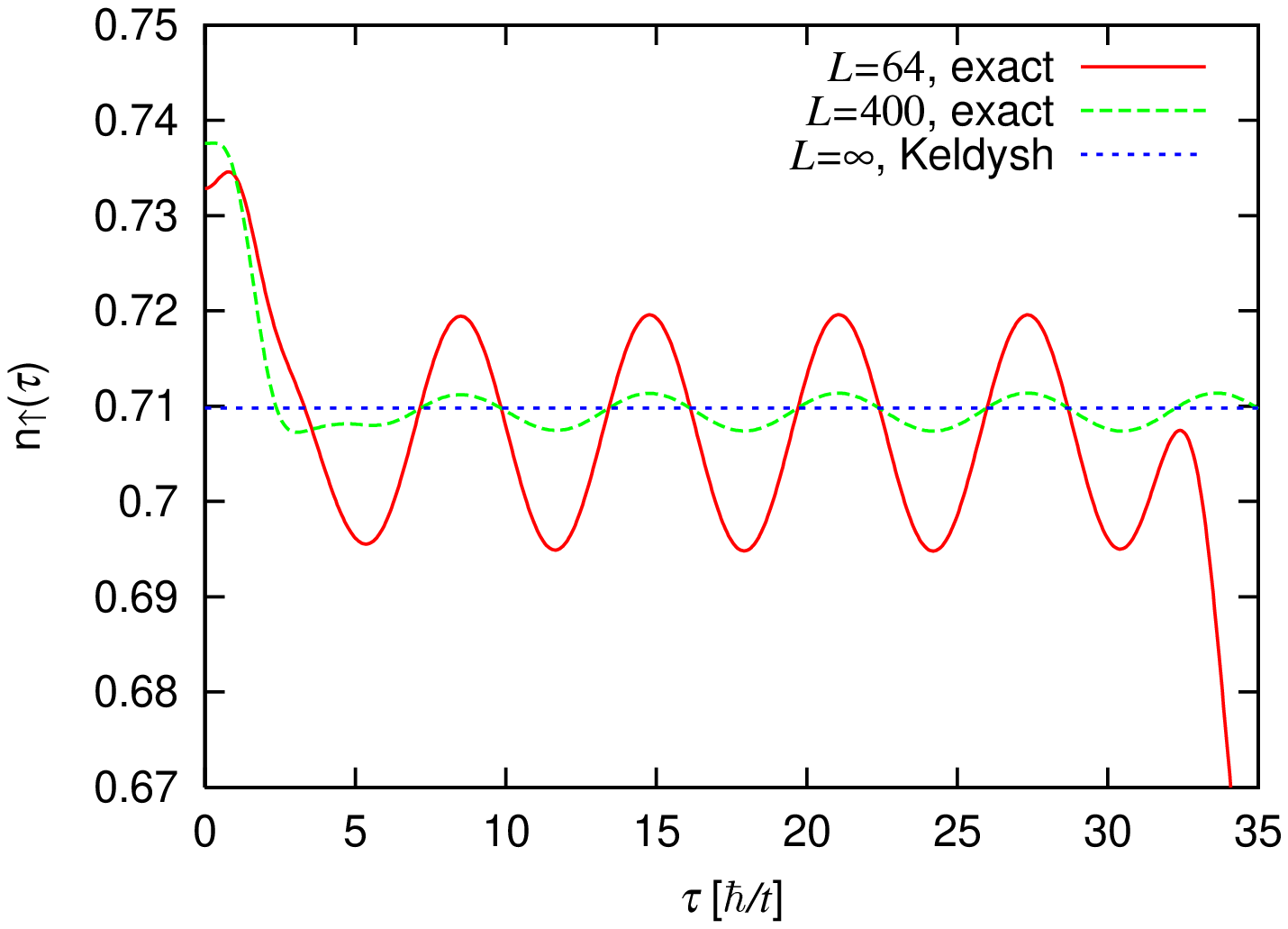}
  \caption{(Color online) $J(\tau)$ (upper figure) and $n_{\uparrow}(\tau)$ (lower
  figure) with different sistem sizes
  obtained by the exact diagonalization ($L=64, L=400$) and analytic
  calculation ($L=\infty$).
  The parameters used are $t'/t=0.6, U/t=0, eV/t=1.0$ and $\tilde h/t=0.5$.
  \label{Fig:1/L_correction}}
 \end{center}
\end{figure}

Thus far we have concentrated on the zero magnetic field case,
where the
system size dependence of the physical quantities is small due to the
symmetric condition.
By making $\tilde h$ finite,
small corrections appear in the values of steady
current $J(V)$ and expectation values of
$n_{\uparrow} \equiv c_{0 \uparrow}^{\dagger} c_{0 \uparrow}$ both in equilibrium
$n_{\uparrow}(\tau=0)$ and in the quasi-steady state $n_{\uparrow}(V)$
as is seen in Fig.\ref{Fig:1/L_correction} for the noninteracting case.
Here the oscillation in $n_{\uparrow}(\tau)$ comes from the oscillations
in $J_L(\tau)$ and $J_R(\tau)$, and we define $n_{\uparrow}(V)$ as the
central value of the oscillation in the quasi-steady state region.
The finite size correction in $J(V)$ has a plus sign and the correction in
$n_{\uparrow}(V)$ has a minus sign.
The system size dependence of these corrections in $J(V)$ and 
$n_{\uparrow}(V)$ is found to be $1/L$.

This phenomena can be explained as follows:
when the spin polarization at the dot $\langle S^z \rangle$ becomes
finite by the symmetry breaking caused by the magnetic field,
the polarization of the lead part as a whole is
$-\langle S^z \rangle$, since $\sum_{i}S_{i}^z$ is fixed to zero.
Because of this finite spin polarization the lead part of a finite
system is shifted from the siglet state, which is the ground state of
the lead part,
whereas for an infinite system $-\langle S^z \rangle$ does not cause any
effect to the lead part.
$\langle S^z \rangle$ of a finite system is determined
energetically by the balance between the Zeeman energy and the energy
difference of the lead part from the siglet state.
The latter is important for relatively short system, suppressing
$\langle S^z \rangle$.
Thus, it can be concluded that the magnetic field is effectively reduced
by the finite size effect.
This effect makes $J(V)$ larger compared to that of longer system.

In the interacting case, it is expected that this finite size effect
becomes stronger than the $U=0$ case because of the enhancement of 
the polarization by the Coulomb repulsion.

Though we can remove this $1/L$ correction by the finite size scaling,
we simply neglect it in this paper.
Hence the corrections are included in the $\tilde h \neq 0$ results in later
discussions, but this is not important for $0 \leq U/t \leq 2$.

\section{Physical Properties of the Steady State}
In this section we discuss the dependence of physical quantities on the
Coulomb interaction $U$, the bias voltage $V$ and the magnetic field
$\tilde h$.
The results presented in this section are obtained as follows:
for each value of $V$ we perform the TdDMRG calculation, get the
quasi-steady state and take the average value over the quasi-steady state region.

\subsection{Zero magnetic field}
\subsubsection{Current \label{Sec:current_0}}
\begin{figure}[t]
 \begin{center}
  \includegraphics[width=4.5cm,angle=270]{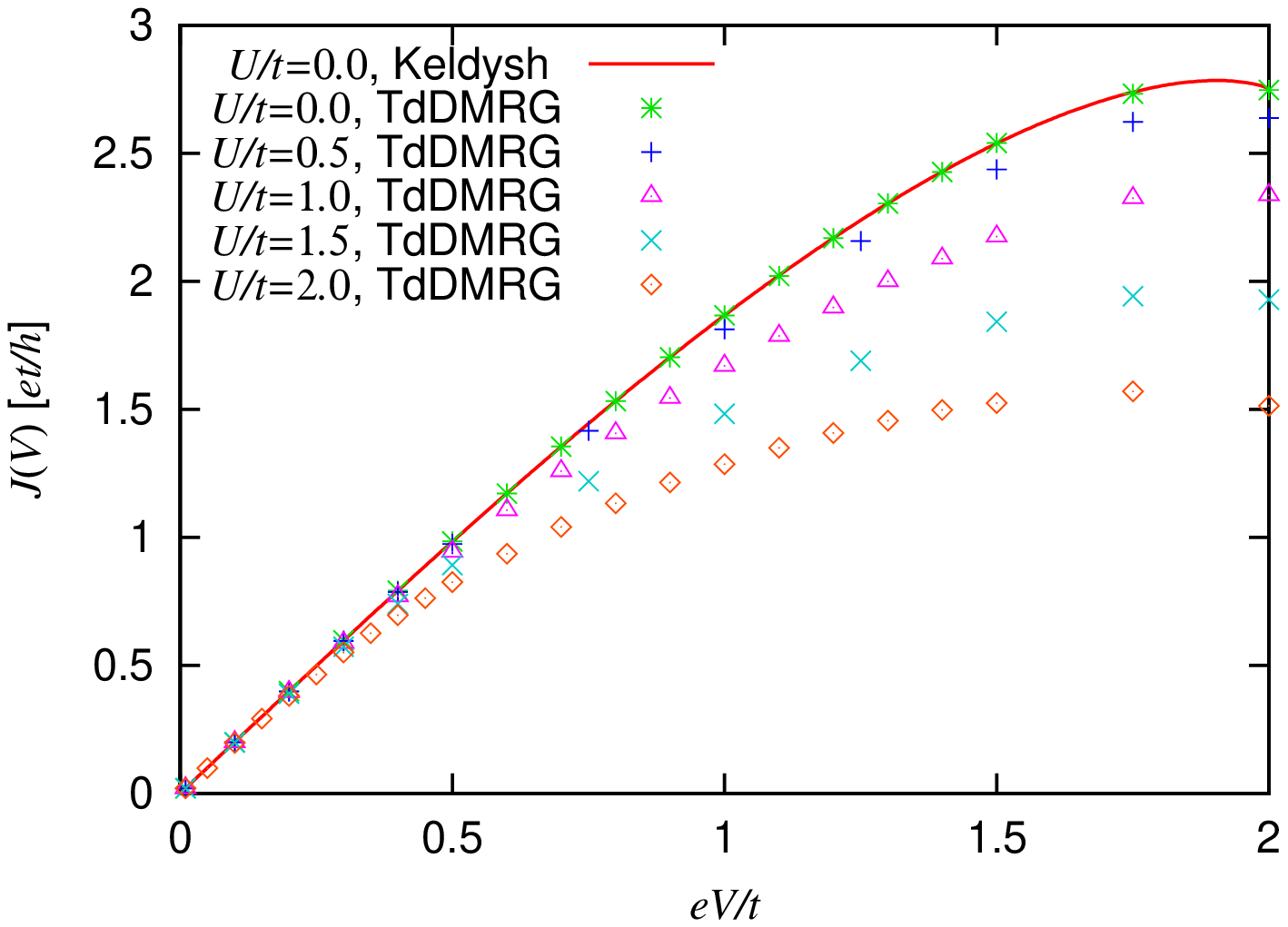}
  \caption{(Color online) The current obtained via eq.\eqref{nonequilibrium_current}
  for $L=\infty, U/t=0$ and the current calculated as the averages over
  the quasi-steady state region of the TdDMRG results.
  The parameters used are $L=64, \tilde h/t=0$, $m=800$ for
  $U/t=2,  eV/t=2$ and $m=600$ for the others.
  \label{Fig:current_Ez=0}}
  \includegraphics[width=4.5cm,angle=270]{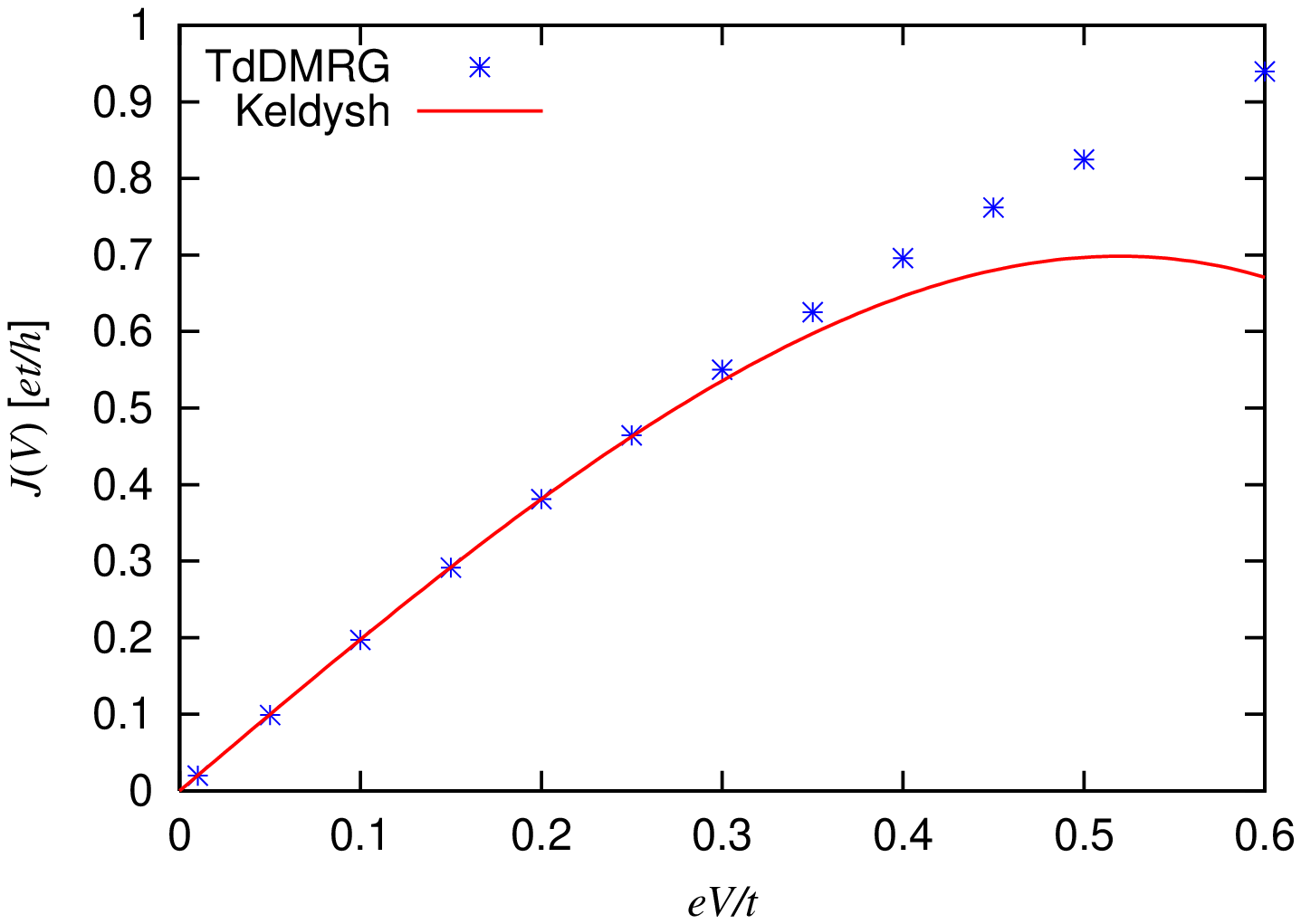}
  \caption{(Color online) The TdDMRG current and $J(V)$ analytically calculated with
  the use of eqs.\eqref{nonequilibrium_current} and \eqref{Oguri_asymptotic}, both for $U/t=2$.
  Note that the latter is asymptotically correct in the low bias regime.
  The TdDMRG result here is the same data set shown in
  Fig.\ref{Fig:current_Ez=0}.
  \label{Fig:low_V}}
 \end{center}
\end{figure}

Fig.\ref{Fig:current_Ez=0} summarizes our results of $J(V)$.

\begin{figure}[t]
 \begin{center}
  \includegraphics[width=4.5cm,angle=270]{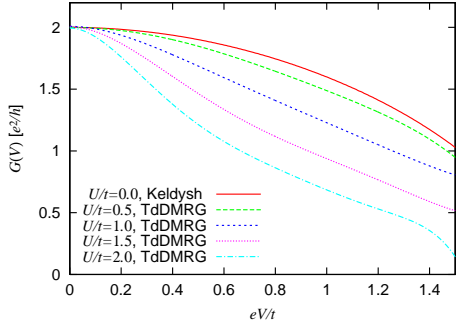}
  \caption{(Color online) The differential conductance for various $U/t$ obtained by
  the integration of eq.\eqref{nonequilibrium_current} ($U/t=0$) and
  the TdDMRG and interpolation.
  The parameters used are the same as in Fig.\ref{Fig:current_Ez=0}.
  \label{Fig:G(V)_Ez=0}}
 \end{center}
\end{figure}%
In the noninteracting case we see nice agreement between the TdDMRG
and the analytic results obtained via eq.\eqref{nonequilibrium_current}, 
for all bias voltages.

For $U\neq 0$ let us compare the TdDMRG results with analytic
ones in the low bias regime.
For this purpose we use an asymptotic expression for $\rho$ in the low
energy, low bias voltage and low temperature \cite{Oguri},
\begin{align}
 \rho_{\sigma}(\omega)=\frac{1}{\pi \Gamma} \bigg[ 1
 &-\left(\tilde \chi_{\uparrow \uparrow}^2 + \frac{1}{2} \tilde
 \chi_{\uparrow \downarrow}^2\right)
 \left(\frac{\omega}{\Gamma}\right)^2
- \frac{1}{2} \tilde \chi_{\uparrow \downarrow}^2
 \left(\frac{\pi T}{\Gamma}\right)^2    \notag \\
 &- \frac{3}{8} \tilde \chi_{\uparrow \downarrow}^2
 \left(\frac{eV}{\Gamma}\right)^2 + \cdots
\bigg].
\label{Oguri_asymptotic}
\end{align}
Putting this expression into eq.\eqref{nonequilibrium_current}, we obtain the
asymptotic behavior of $J(V)$ for small $V$.
From the comparison for $U/t=2$, Fig.\ref{Fig:low_V},
we see that the TdDMRG result is in excellent accordance with the analytic one at
$eV/t \leq 0.25$.
Therefore we can conclude that our TdDMRG calculation and definition of the
quasi-steady states work well for $U/t \leq 2$ at least for low $V$.

Back to Fig.\ref{Fig:current_Ez=0}, in the high bias voltage we see 
$J(V=1.75t/e) > J(V=2t/e)$.
This behavior is due to the energy dependence of $\Gamma_{L,R}$, thus
representing a specific nature of the 1-D nearest-neighbor tight-binding leads.
Since we are interested in the universal behavior of the Kondo effect in
the quantum dot system, we will restrict ourselves to study up to $eV/t=2$.

\subsubsection{Differential conductance}
In many works on the nonequilibrium transport phenomena,
the differential conductance
$G(V) \equiv \frac{\partial J(V)}{\partial V}$ has been discussed as an
extention of the linear conductance.

Here we carry out the following process to get $G(V)$ from the data
points of $J(V)$.
We interpolate the $J(V)$ for $0 \leq eV/t \leq 1.5$ by the least square
fitting to an odd polynomial expression.
Then we differentiate the resulting function and obtain
$G(V)$, Fig.\ref{Fig:G(V)_Ez=0}.
Note that $G(V)$ calculated in this way contains errors due to the
interpolation especially near $eV/t=1.5$, the edge part of the data points.

In Fig.\ref{Fig:G(V)_Ez=0} the unitarity limit of the zero bias peak
$G(0) = 2e^2/h$ are clearly seen for every $U$.
$G(V)$ drops from the zero bias peak with increasing $V$.
The width of the zero bias peak becomes narrower as $U$ increases.
These behaviors result from the well known fact that 
the width of the Kondo peak in $\rho_{\sigma}(\omega)$ is essentially
given by the Kondo temperature $T_K$, and that 
the peak height of the Kondo peak is suppressed by the bias voltage. 
Fig.\ref{Fig:G(V)_Ez=0} demonstrates that the TdDMRG succesfully
reproduces the characteristic properties of $G(V)$ \cite{2nd_order, NCA, Fujii_1}.

\subsection{Effect of magnetic field}
\subsubsection{Current and differential conductance}
\begin{figure}[t]
 \begin{center}
  \includegraphics[width=4.5cm,angle=270]{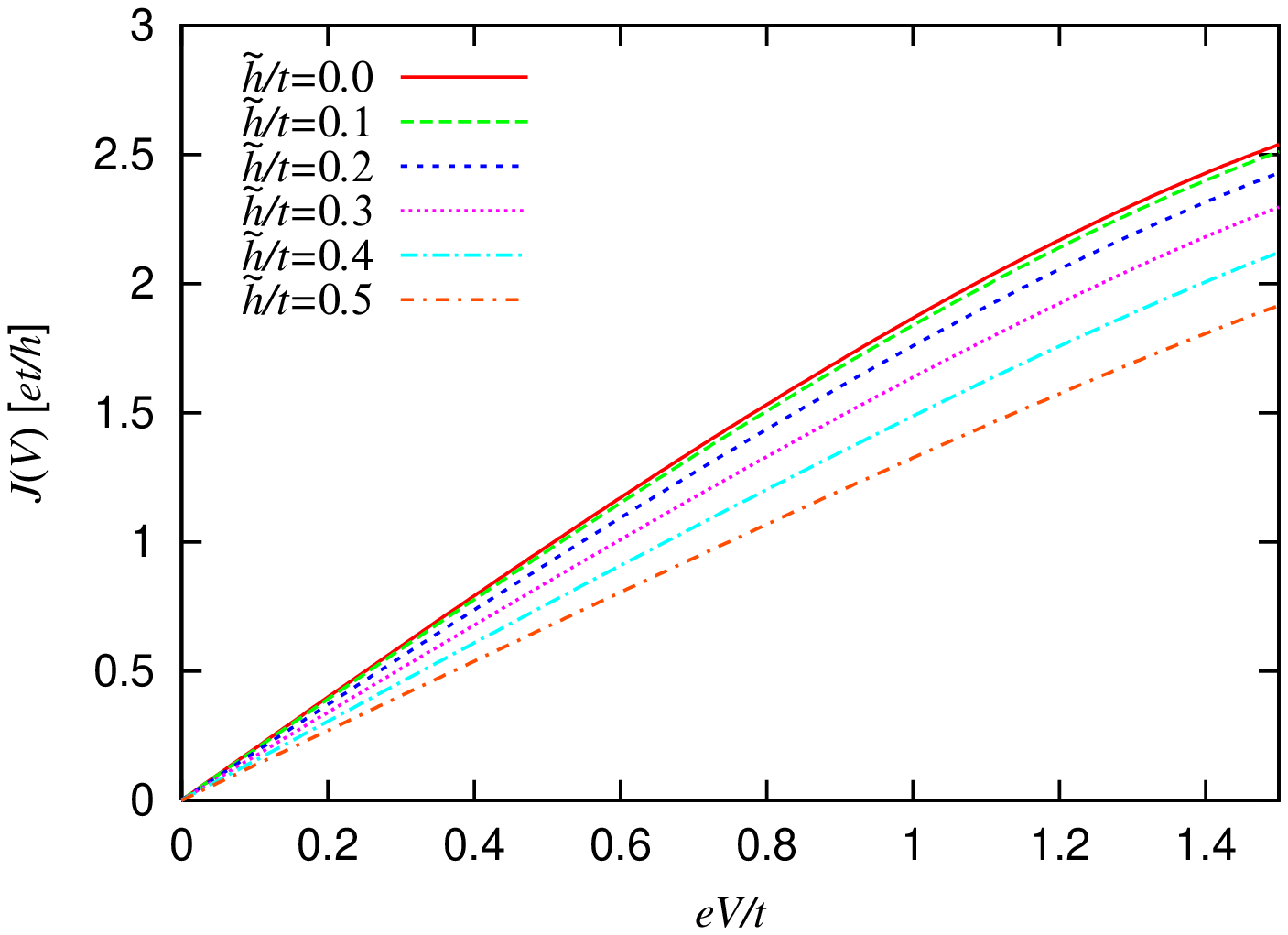}
  \includegraphics[width=4.5cm,angle=270]{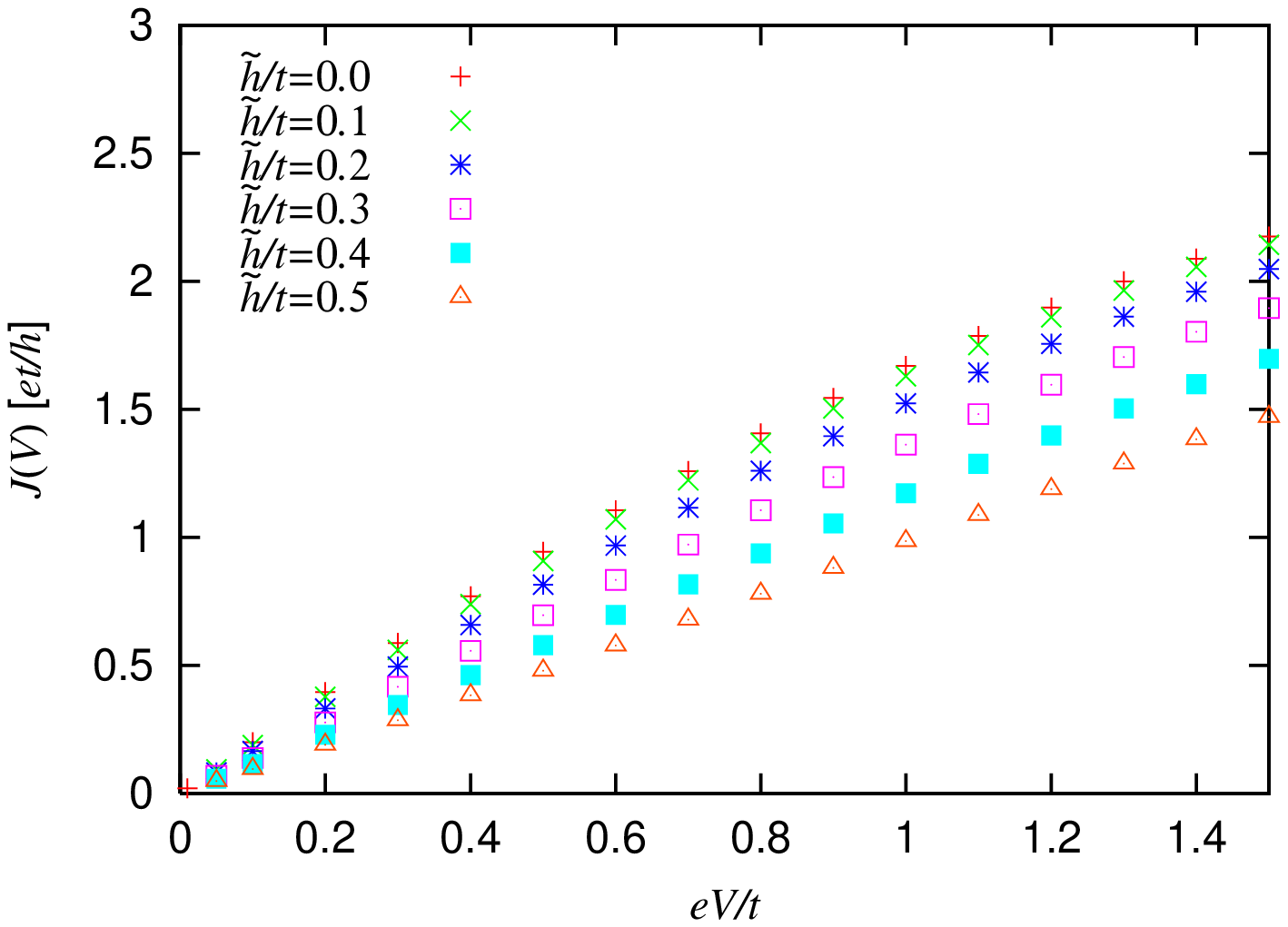}
  \includegraphics[width=4.5cm,angle=270]{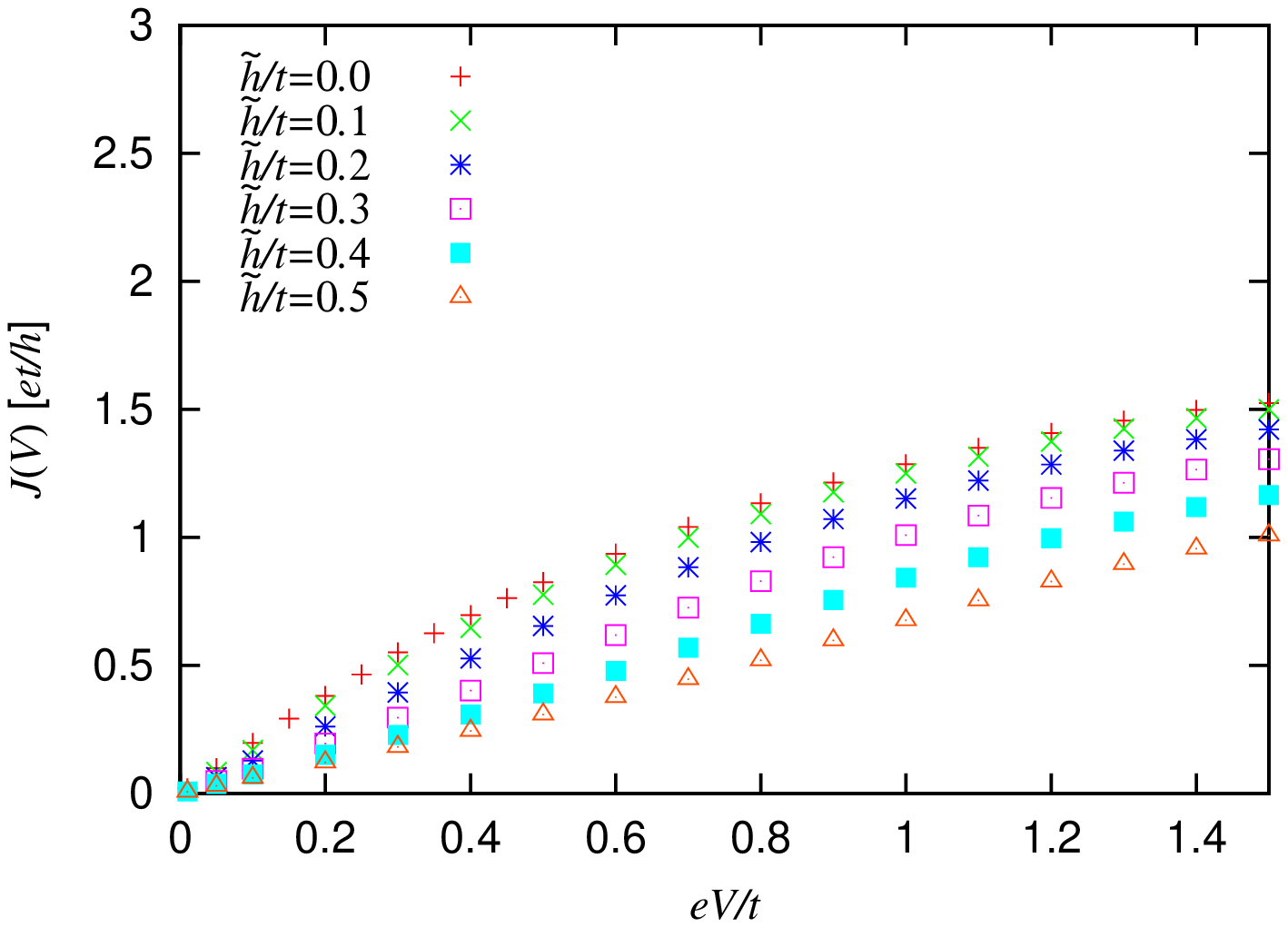}
  \caption{(Color online) The current under a magnetic field for $U/t=0,1,2$ from the top.
  The top ($U/t=0$) result is obtained by the integration of
  eq.\eqref{nonequilibrium_current} and the others are obtained by the
  TdDMRG calculation.
  The results for $\tilde h/t=0$ are the same ones in Fig.\ref{Fig:current_Ez=0}.
  The other TdDMRG results are obtained with $L=64, m=1024$ and the OBC.
  \label{Fig:J(V)_Ez!=0}}
 \end{center}
\end{figure}

\begin{figure}[t]
 \begin{center}
  \includegraphics[width=4.5cm,angle=270]{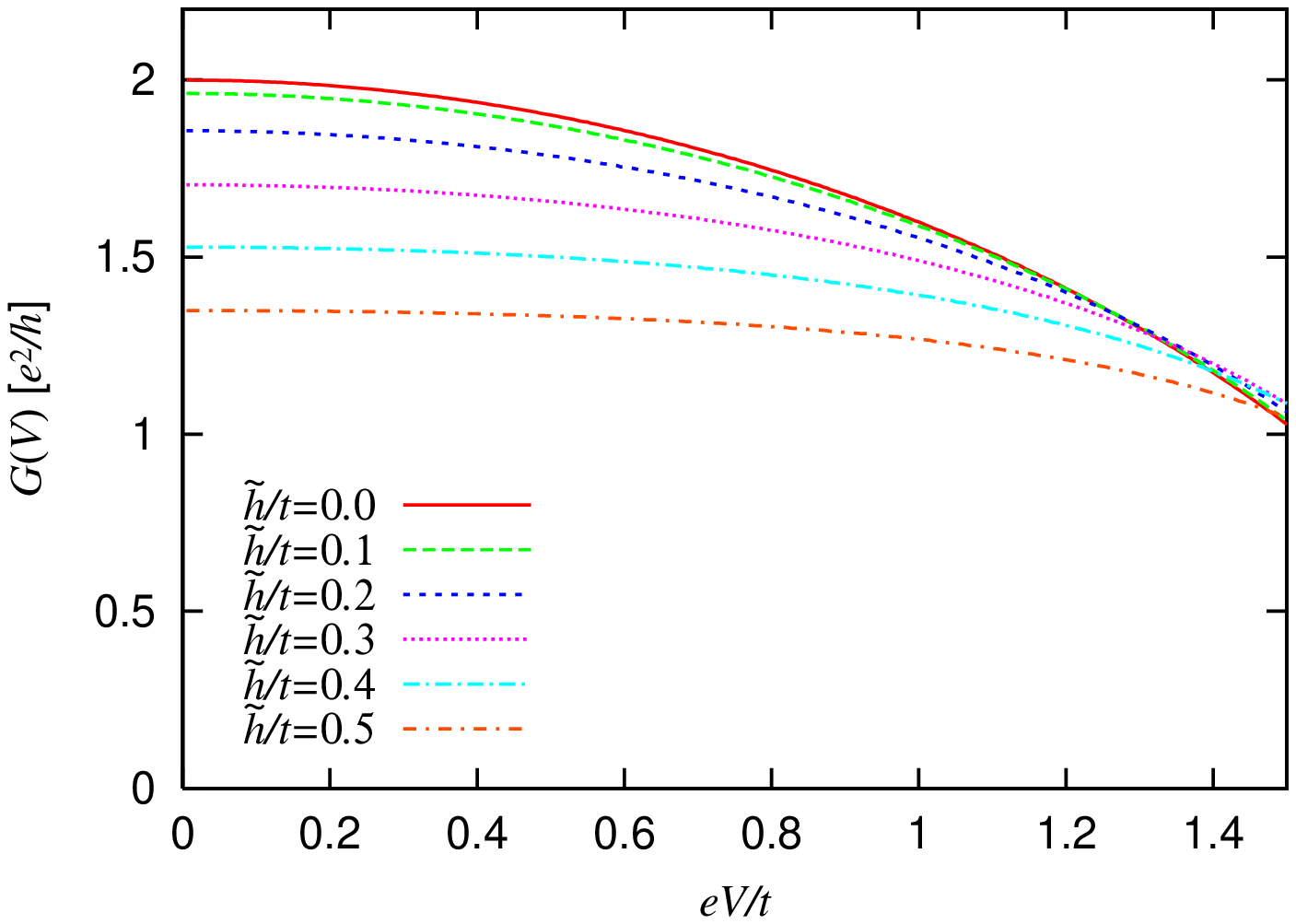}
  \includegraphics[width=4.5cm,angle=270]{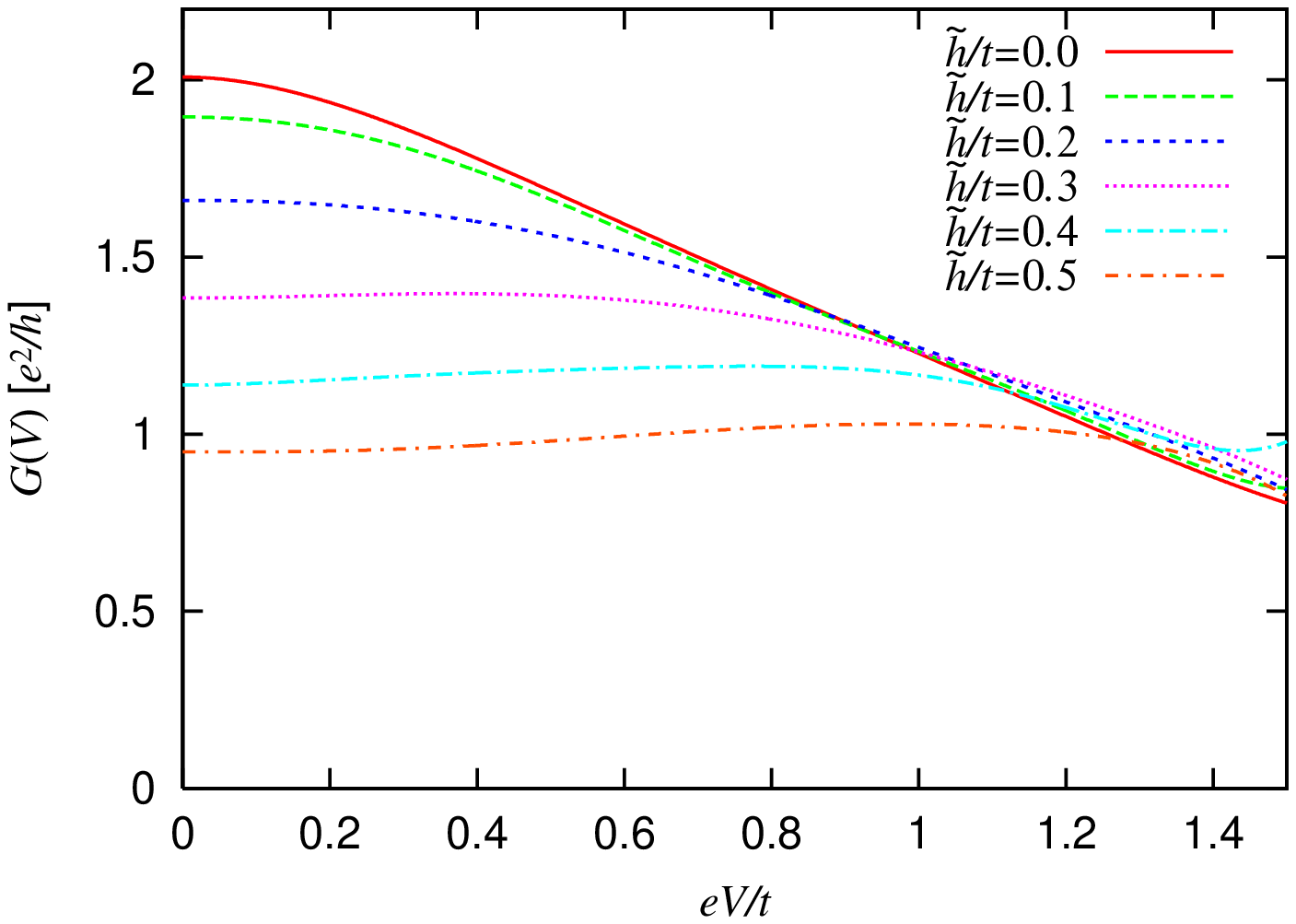}
  \includegraphics[width=4.5cm,angle=270]{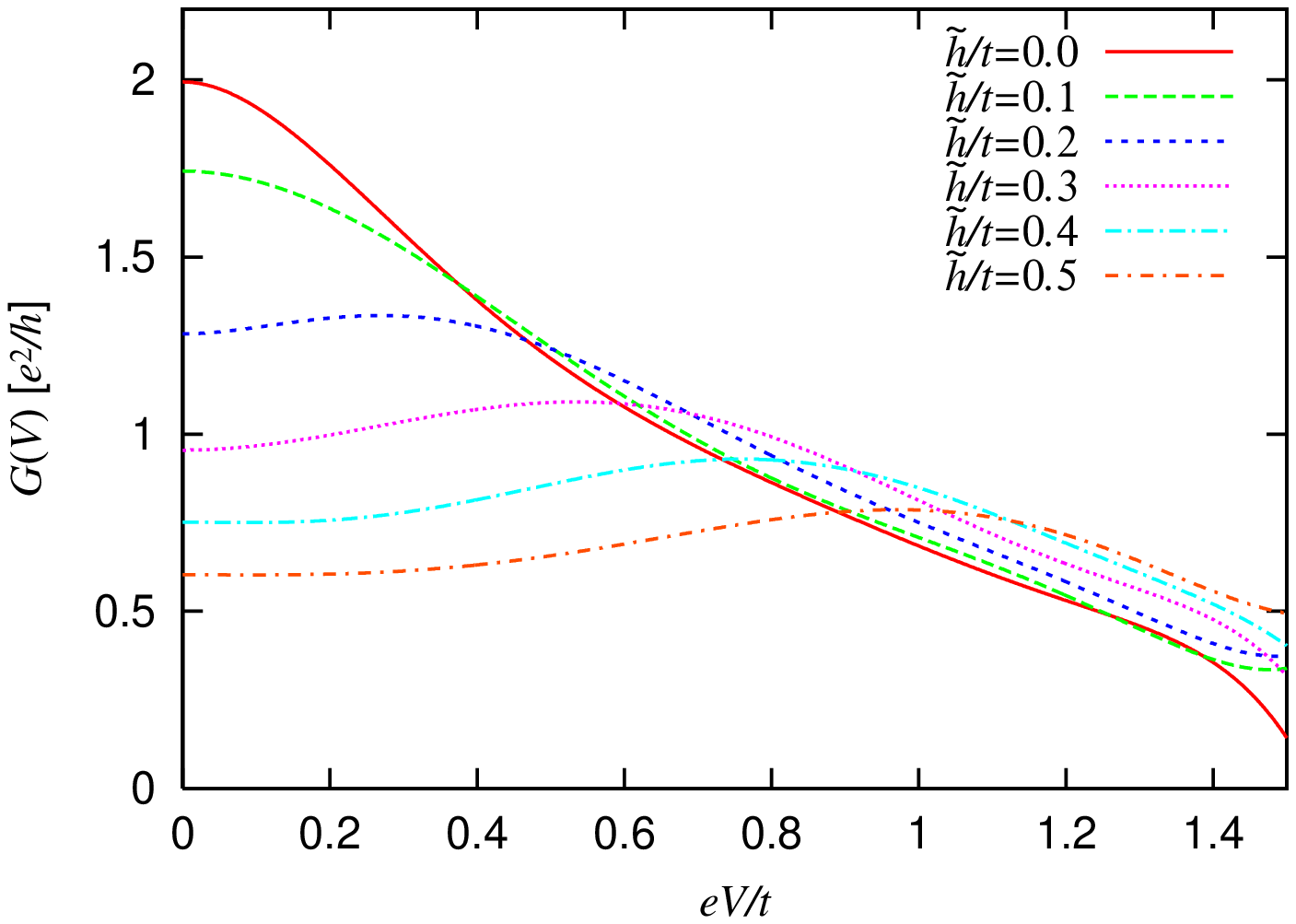}
  \caption{(Color online) The differential conductance under a magnetic field for
  $U/t=0,1,2$ from the top, obtained from the data shown in
  Fig.\ref{Fig:J(V)_Ez!=0}.
  \label{Fig:G(V)_Ez!=0}}
 \end{center}
\end{figure}

Even under a finite magnetic field up to $\tilde h/t \leq 0.5$ our
TdDMRG calculation does not lose its validity to study the steady state,
in spite of the $1/L$ correction discussed in \S \ref{Sec:1/L}.
Using the same definition of the quasi-steady state and the
same interpolation as in the $\tilde h=0$ case,
we get the results for $J(V)$, Fig.\ref{Fig:J(V)_Ez!=0} and $G(V)$, 
Fig.\ref{Fig:G(V)_Ez!=0}.

It is seen that $J(V)$ is suppressed by the magnetic
field for all $U$ and $V$.
Then we observe the splitting of the zero bias peak in $G(V)$ by a
finite magnetic field when $U$ and $\tilde h$ are relatively large.
Moreover the position of the splitted peak is roughly equal to the twice
of the Zeeman energy, $eV = 2 \tilde h$.
These behaviors are consistent with the previous results \cite{NCA,Fujii_2},
and can be naively explained by the width of the peak of
$\rho_{\sigma}(\omega)$ and the separation of $\rho_{\uparrow}(\omega)$
and $\rho_{\downarrow}(\omega)$ by the Zeeman energy.
When $U$ is large the peak is narrow, then the separated peak is not
included in the interval of integration in eq.\eqref{nonequilibrium_current},
if $eV$ is small compared to the Zeeman energy.
Thus in this case $G(V)$ has the splitted peaks corresponding to the peaks of
$\rho_{\sigma}(\omega)$.
On the other hand, when $U$ is small the separation of
the broad peaks in $\rho_{\sigma}(\omega)$ does not affect the peak
structure of $G(V)$.

Again we can conclude that our TdDMRG calculation works properly
for the calculation of $G(V)$ in magnetic fields.

\subsubsection{Magnetic properties}
\begin{figure}[t]
 \begin{center}
  \includegraphics[width=4.5cm,angle=270]{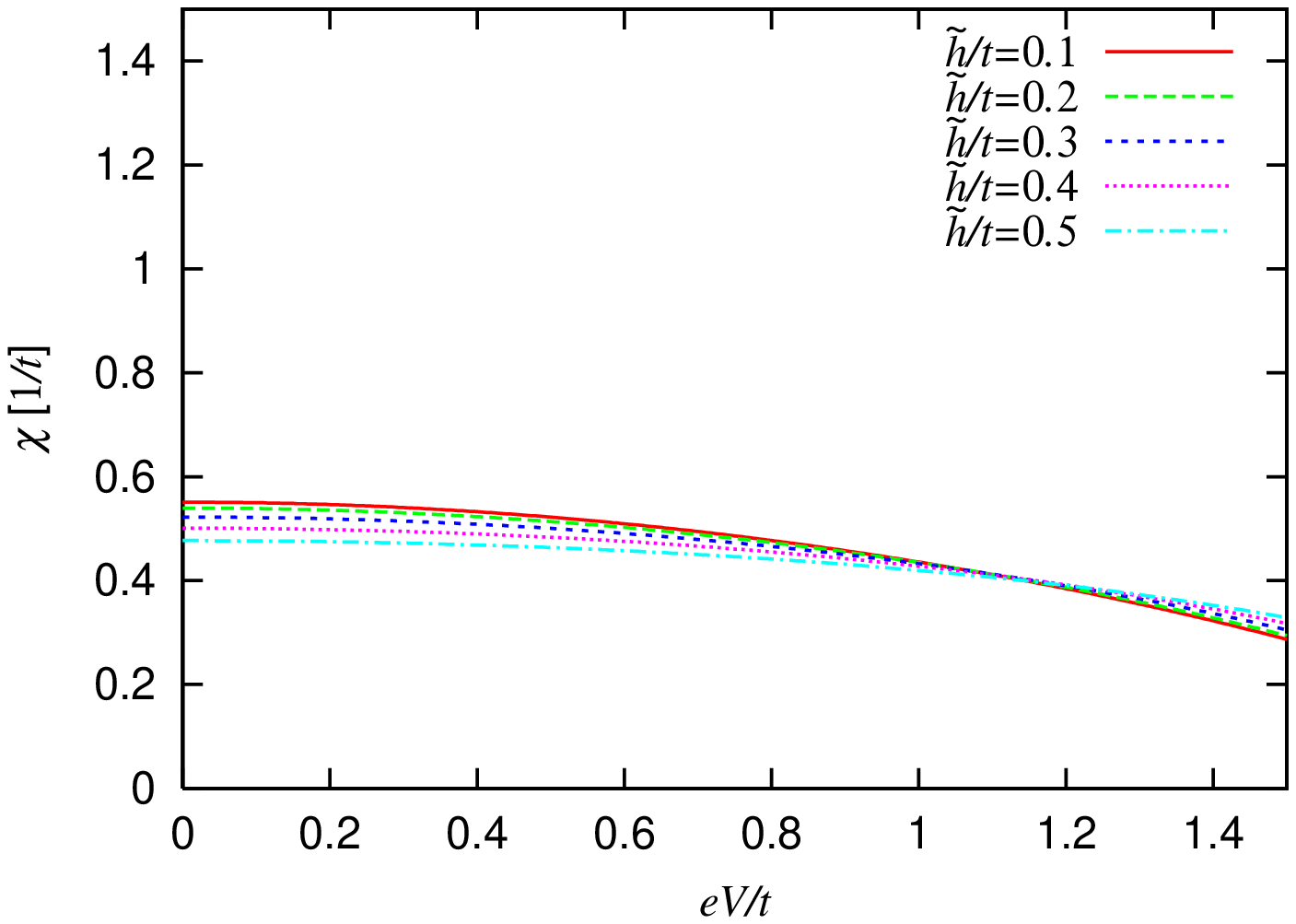}
  \includegraphics[width=4.5cm,angle=270]{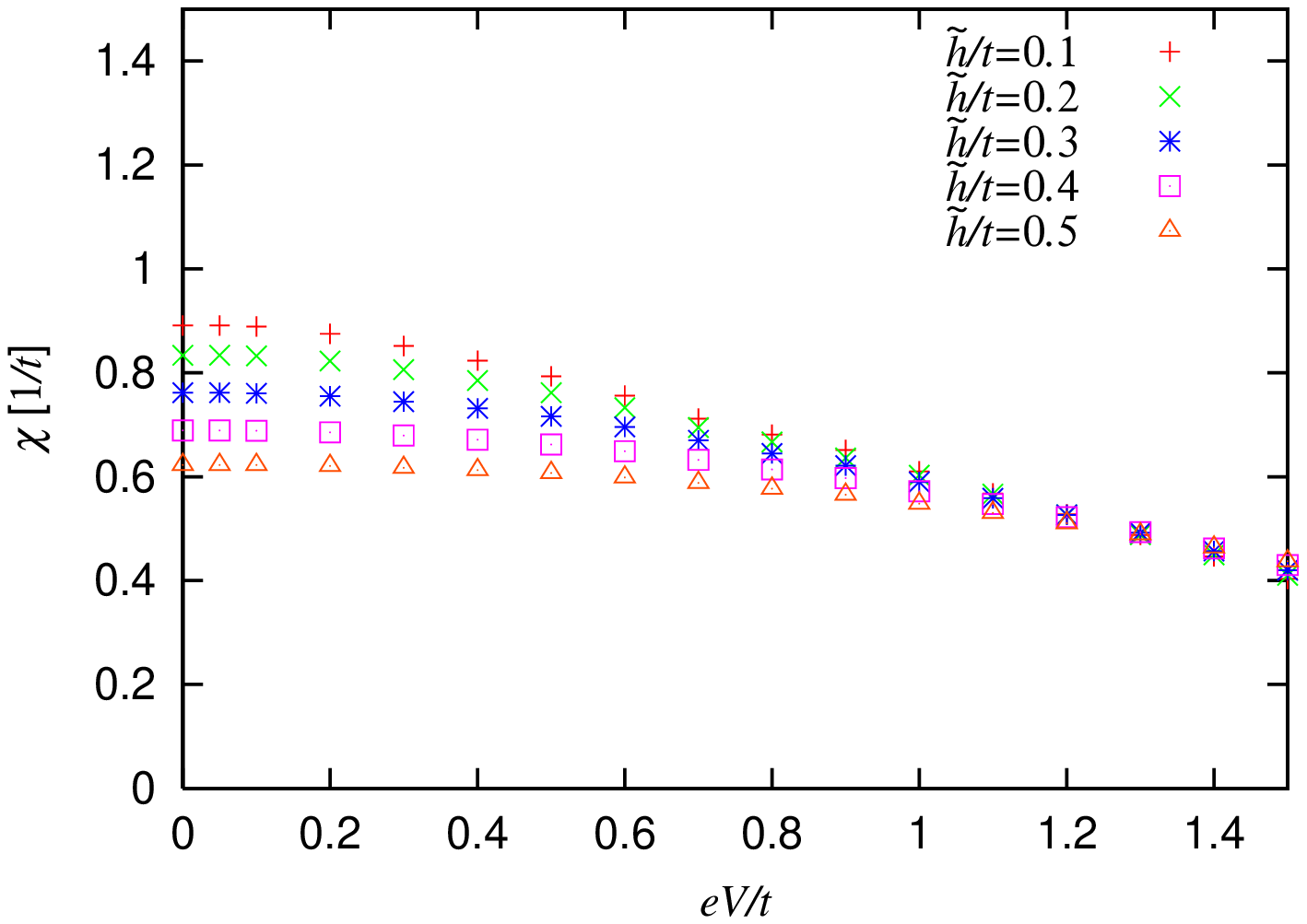}
  \includegraphics[width=4.5cm,angle=270]{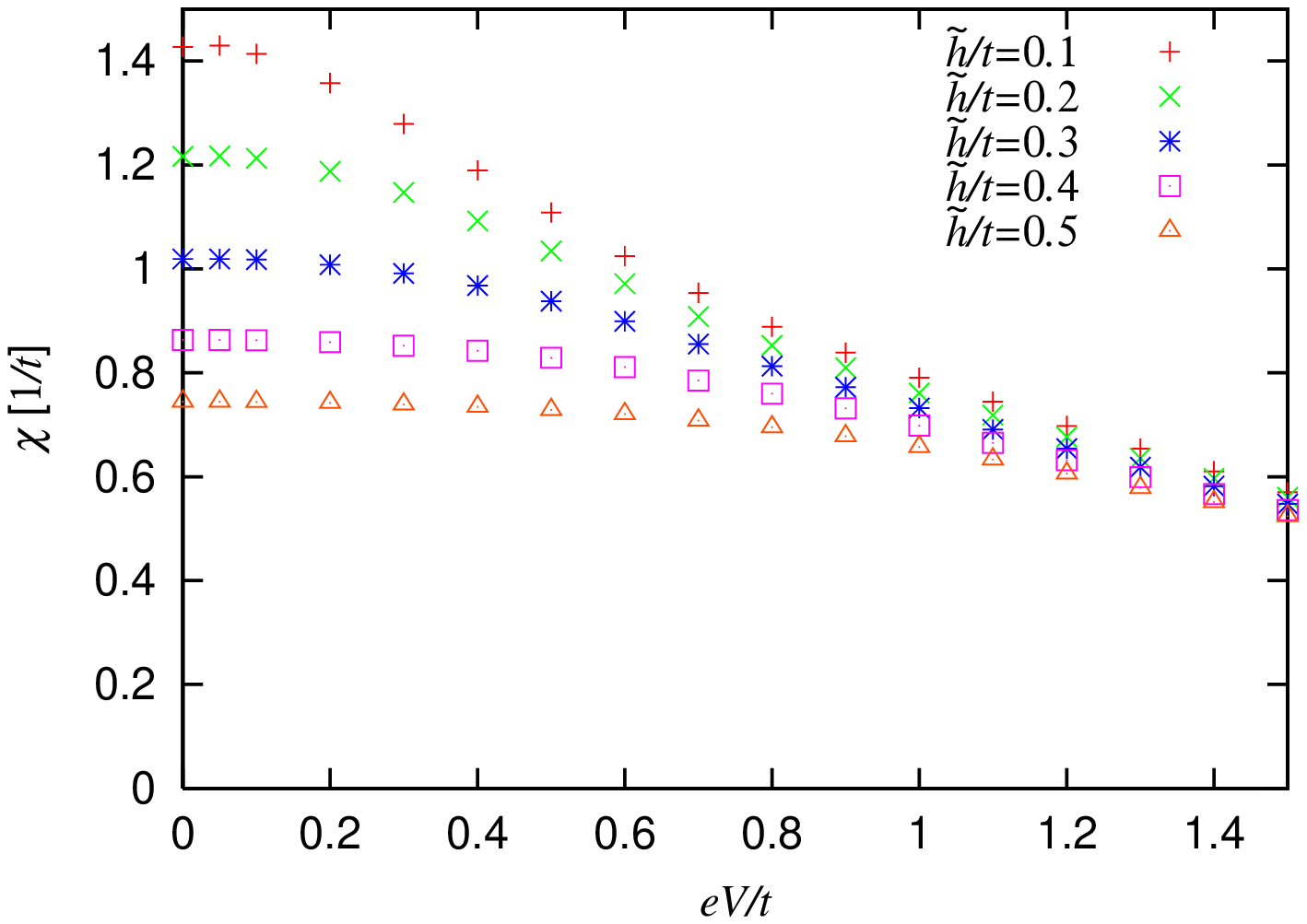}
  \caption{(Color online) The magnetic susceptibility defined as in eq.\eqref{susceptibility}
  for $U/t=0,1,2$ from the top.
  The figure at the top ($U/t=0$) is obtained by
  eq.\eqref{nonequilibrium_n}.
  The others are obtained by the TdDMRG with $L=64, m=1024$ and the OBC.
  \label{Fig:chi(V)}}
 \end{center}
\end{figure}

With the knowledge of $| \psi(\tau) \rangle$, of course, one can take expectation
values of various operators other than the current operator.
In this paper we present here the susceptibility, Fig.\ref{Fig:chi(V)}, which we define as
\begin{align}
 \chi(V) \equiv \frac{\langle S^{z} \rangle}{\tilde h},
\label{susceptibility}
\end{align}
where magnetization at the dot $\langle S^{z} \rangle$ is the
averaged value of $\langle \psi (\tau) | S^{z} | \psi(\tau) \rangle$
over the quasi-steady state region.
The numerically calculated 
$\langle \psi (\tau) | n_{\uparrow} | \psi(\tau) \rangle$
shows the oscillation as in the lower figure of Fig.\ref{Fig:1/L_correction},
but the oscillation can be removed by subtracting 
$\langle \psi (\tau) | n_{\downarrow} | \psi(\tau) \rangle$.
In Keldysh formalism the magnetization can be calculated from
\begin{align}
 \langle n_{\sigma} \rangle = \langle c_{0 \sigma}^{\dagger} c_{0
 \sigma} \rangle
 = -i \int_{-\infty}^{\infty} \frac{d \omega}{2 \pi}
 G_{\sigma}^{-+}(\omega),
\label{nonequilibrium_n}
\end{align}
where $G_{\sigma}^{-+}(\omega)$ is the Fourier component of one of the Keldysh Green functions,
$G_{\sigma}^{-+}(\tau) \equiv i \langle c_{0\sigma}^{\dagger}(0) c_{0 \sigma}(\tau) \rangle$.

Let us start the interpretation of Fig.\ref{Fig:chi(V)} from $V=0$.
It is well known from the analyses of the Kondo effect in equilibrium
that the Coulomb repulsion $U$ enhances the linear susceptibility
$\lim_{\tilde h \rightarrow \infty} \chi(V=0)$
and a finite magnetic field suppresses $\chi(V=0)$.
We can clearly see these behaviors in Fig.\ref{Fig:chi(V)}.

Next the effect of $V$ is explained qualitatively by considering the
simplest case, where $U=0$ and the density of states of the leads
is constant.
Then we easily find
\begin{align}
 \langle S^z \rangle = \frac{1}{2} \int_{-\infty}^{-eV/2} d \omega \left(
 \rho_{\uparrow}(\omega) - \rho_{\downarrow}(\omega) \right),
\end{align}
using the symmetric condition.
From this expression we see that
when $\tilde h \gg eV$ the peak of $\rho_{\uparrow}$ is included in the
integration but not the peak of $\rho_{\downarrow}$, resulting in a large
value of $\langle S^z \rangle$.
On the other hand,
$\rho_{\uparrow}$ and $\rho_{\downarrow}$ are both small in the interval of
integration when $\tilde h \ll eV$.
Thus we can expect that large $V$ suppresses the spin polarization and the
susceptibility, as can be seen in Fig.\ref{Fig:chi(V)}.
Even when $U\neq 0$ and the density of states has energy dependence,
the above story roughly holds and explains the $V$ dependence of
$\chi(V)$ in Fig.\ref{Fig:chi(V)}.

\subsection{Behavior of $\langle n_{\uparrow} n_{\downarrow} \rangle$}
In this subsection let us discuss the $U,V$ and $\tilde h$ dependence of
$\langle n_{\uparrow} n_{\downarrow} \rangle$.
Since this operator appears as $U n_{\uparrow} n_{\downarrow}$ in the
Hamiltonian \eqref{Hamiltonian_all}, its expectation value
reflects the effect of electron correlation.

Again $\langle \psi(\tau) | n_{\uparrow} n_{\downarrow} |\psi(\tau) \rangle$
as a function of time shows an oscillation as in $J_{L,R}(\tau)$ and
$n_{\uparrow, \downarrow}(\tau)$.
In this case we remove the oscillation 
by taking the average value of
$\langle \delta n_{\uparrow} \delta n_{\downarrow} \rangle$
over the quasi-steady state region,
where $\delta n_{\sigma} \equiv n_{\sigma} - \langle n_{\sigma} \rangle$.

\begin{figure}[t]
 \begin{center}
  \includegraphics[width=4.5cm,angle=270]{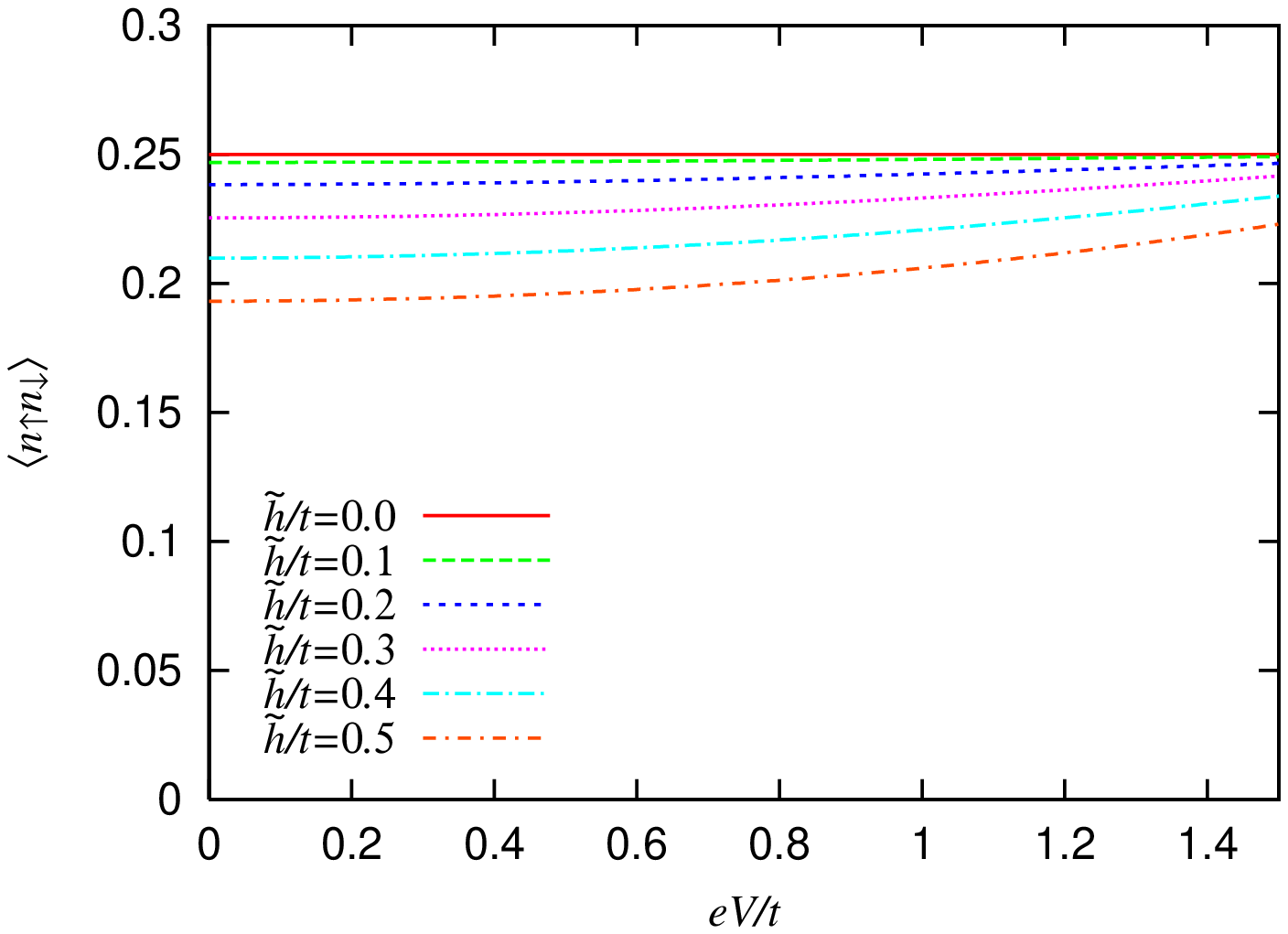}
  \includegraphics[width=4.5cm,angle=270]{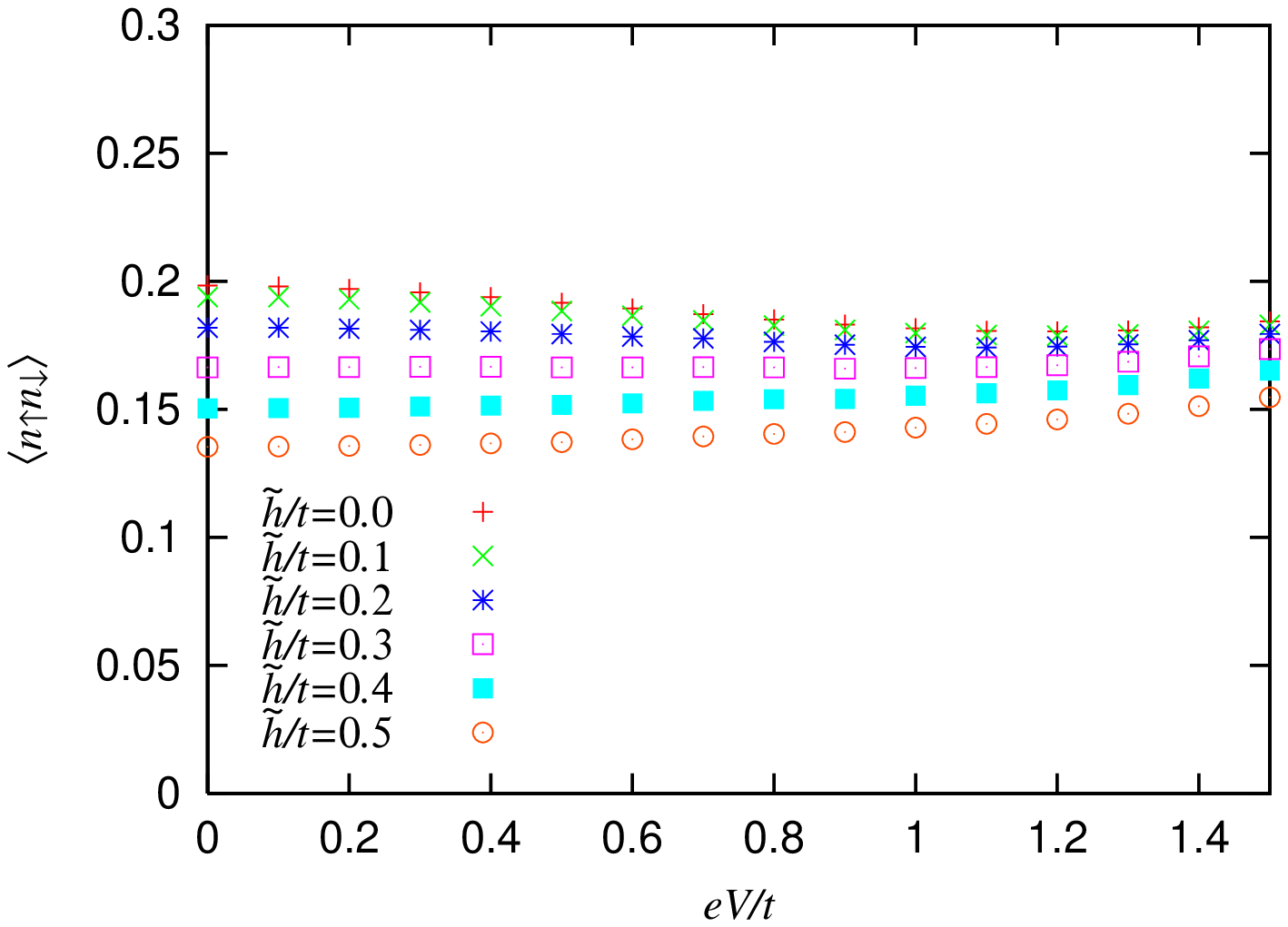}
  \includegraphics[width=4.5cm,angle=270]{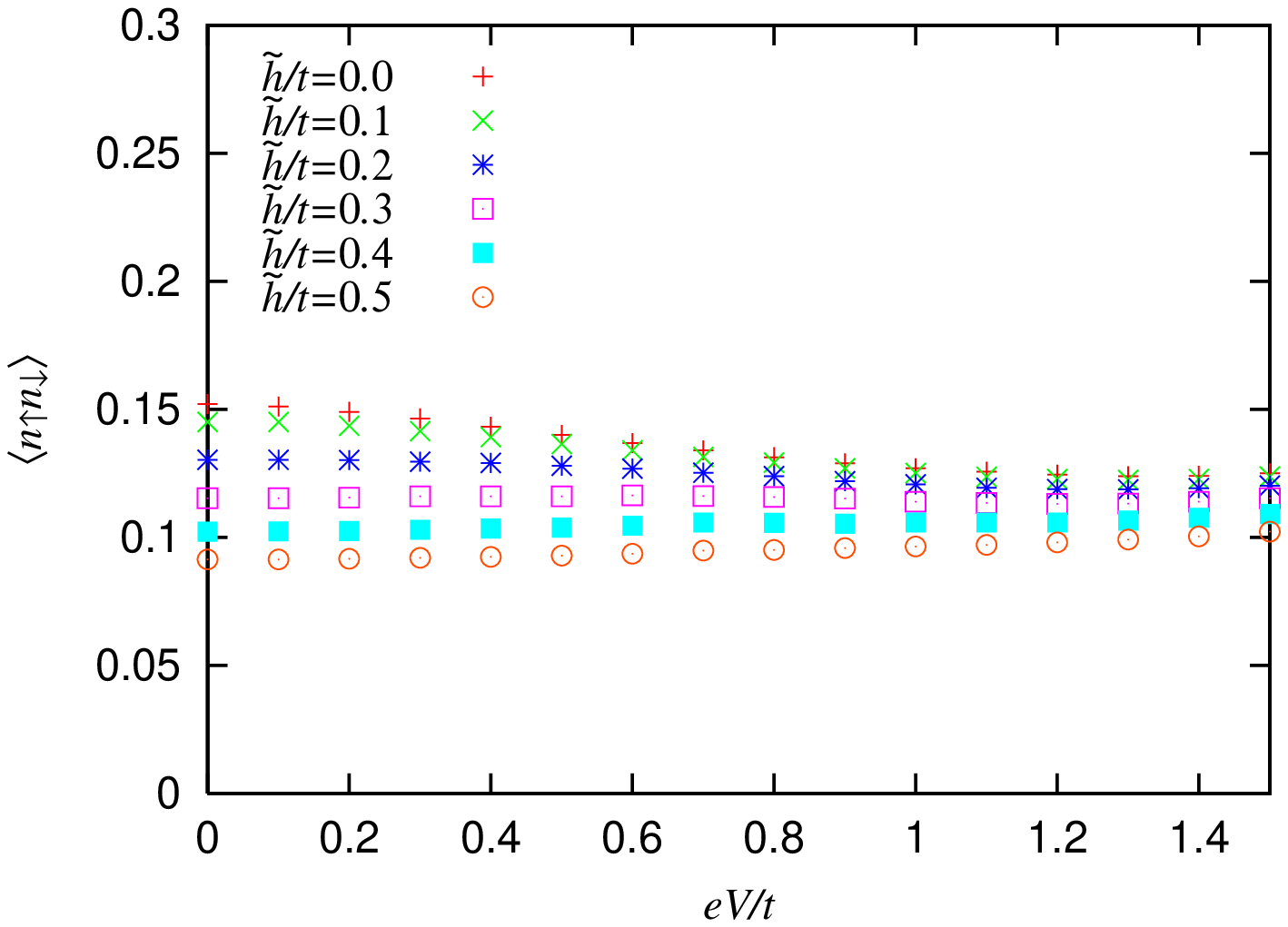}
  \caption{(Color online) The expectation values of $n_{\uparrow} n_{\downarrow}$ taken
  by the quasi-steady state wave function for $U/t=0,1,2$ from the top.
  The top figure ($U/t=0$) is obtained by \eqref{nonequilibrium_n} and 
  $\langle n_{\uparrow} n_{\downarrow} \rangle = \langle n_{\uparrow}
  \rangle \langle n_{\downarrow} \rangle$.
  The others are obtained by the TdDMRG calculation and the parameters
  used are the same ones as Fig.\ref{Fig:chi(V)}.
  \label{Fig:n_up_n_down}}
 \end{center}
\end{figure}

From the previous study it is known that in the high voltage limit
the third-order contributions to the self energies vanish, resulting
in $n_{\sigma} \rightarrow 1/2$ \cite{Fujii_2}.
This is accounted for by effective suppression of electron
correlation and magnetic field by the bias voltage.
Thus it is likely that
$\langle n_{\uparrow} n_{\downarrow} \rangle \rightarrow 1/4$
in the limit of $V \rightarrow \infty$, independent of $\tilde h$.
In Fig.\ref{Fig:n_up_n_down} we see for $U/t=0$,
$\langle n_{\uparrow} n_{\downarrow} \rangle$ approaches to $1/4$
monotonically with increasing $V$.
However, for $U/t =1,2$ there appears nonmonotonic behavior for
relatively small $\tilde h$, while the magnetization is a monotonically
decreasing function of $V$, as can be seen in Fig.\ref{Fig:chi(V)}.
In other words, $\langle n_{\uparrow} n_{\downarrow} \rangle$ has a minimal
value at $eV/t \sim 1.2$.
Consequently the electron correlation seems to be
enhanced effectively by the bias voltage, compared to the equilibrium state.
This behavior is new and needs an explanation.

This anomalous behavior is seen in relatively low voltage regime $eV/t <1$ and
the effect of the energy dependence of $D_{L,R \sigma}$, the local
density of states of the leads, becomes important in the higher bias
voltage regime.
Therefore we may expect that this new behavior is a universal behavior of
the Kondo effect of the nonequilibrium steady state in the quantum dot system.

\section{Conclusions}
In this paper the adaptive time-dependent DMRG method was applied to the
1-D Anderson model with time dependent bias term.
We have shown that the TdDMRG works well for the studies of the quantum
dot out of equilibrium.

In \S \ref{Sec:4} it was seen that from $J(\tau)$ for
$0\leq U/t\leq 2,\,\,\,\, 0\leq eV/t \leq 2,\,\,\,\, 0\leq \tilde h/t \leq 0.5$
obtained by the TdDMRG calculation the quasi-steady states were found.
In order to realize the well-defined quasi-steady states, we made some
technical improvements compared to the previous study \cite{dot+tdDMRG}:
first, we simply made the number of the DMRG basis $m$ larger to keep the accuracy
until sufficiently long quasi-steady state appeared in $J(\tau)$.
Second, we used the slowly changing bias voltage other than the suddenly
changing one, and reduced the effects of the overshoots and the damping
oscillations in $J(\tau)$. 
Third, for low bias regime the SBC was used to obtain flat $J(\tau)$.
By these improvements, the studies of the properties of the
quasi-steady states became possible.

Then we verified that the physical quantities averaged over the
quasi-steady state regions behave exactly like the ones of the real
nonequilibrium steady states in the infinite system, 
except for the $1/L$ corrections for $\tilde h \neq 0$.
The TdDMRG calculation successfully
reproduced the established results for the differential conductance, such
as the zero bias peak $G(0)=2e^2/h$ for $\tilde h=0$,
and the splitting of the zero bias peak caused by a magnetic field
with the peak position $eV=2\tilde h$.
Also for $\chi(V)$ we obtained reasonable results.

Additionally we observed a nonmonotonic behavior in
$\langle n_{\uparrow} n_{\downarrow} \rangle$ as a function of $V$.
We discussed that this may be a universal behavior of
the Kondo effect in quantum dot system.

It is concluded that the TdDMRG method is a useful numerical tool to
investigate the nonequilibrium steady states of quantum dot systems.
However it should be noted that there is a restriction to this method due to the large Kondo
cloud in the strong coupling regime. 
In order to correctly describe the Kondo physics in the strong coupling
regime, one must employ a very long system and therefore one has to 
take a very large value of $m$.
In this paper we studied relatively weak coupling regime $U/t \leq 2$
and use the SBC to control this size effect.
For larger $U$ it is expected that much more time is required to
correctly realize the quasi-steady states especially in the low bias
regime.
Except for this limitation, the method can be applied to other
nonequilibrium problems of mesoscopic systems.

\begin{acknowledgments}
This research was partially supported by 
the Ministry of Education, Science, Sports 
and Culture, Grant-in-Aid for Young Scientists (B), 17740187, 2005.
 J. Zhao is supported by the Japan Society for the Promotion of Science (P07036).
\end{acknowledgments}

\end{document}